\documentclass[runningheads,fleqn]{llncs}

\usepackage{latexsym}
\usepackage{amssymb}
\usepackage{amsmath}
\usepackage[all]{xy}
\usepackage{hyperref}
\usepackage{wrapfig}
\usepackage{graphicx}

\usepackage{listings}% http://ctan.org/pkg/listings
\newcommand{\ignore}[1]{} 

\newcommand{\id}{{\mathit{id}}} %% empty substitution

\newcommand{\sleq}{\leqslant}
\def\tto{\twoheadrightarrow}

\def\defemb#1#2{\expandafter\def\csname #1\endcsname
                              {\relax\ifmmode #2\else\hbox{$#2$}\fi}}
                              
\defemb{cA}{{\cal A}}
\defemb{cB}{{\cal B}}
\defemb{cC}{{\cal C}}
\defemb{cD}{{\cal D}}
\defemb{cE}{{\cal E}}
\defemb{cF}{{\cal F}}
\defemb{cG}{{\cal G}}
\defemb{cH}{{\cal H}}
\defemb{cI}{{\cal I}}
\defemb{cJ}{{\cal J}}
\defemb{cL}{{\cal L}}
\defemb{cM}{{\cal M}}
\defemb{cN}{{\cal N}}
\defemb{cO}{{\cal O}}
\defemb{cP}{{\cal P}}
\defemb{cQ}{{\cal Q}}
\defemb{cR}{{\cal R}}
\defemb{cS}{{\cal S}}
\defemb{cT}{{\cal T}}
\defemb{cU}{{\cal U}}
\defemb{cV}{{\cal V}}
\defemb{cW}{{\cal W}}
\defemb{cX}{{\cal X}}
\defemb{cZ}{{\cal Z}}

\newcommand{\pos}{{\cP}os}

\newcommand{\Var}{\mathsf{{\cal V}ar}} % variables in an object

\newcommand{\dom}{{\cD}om}

\newcommand{\cons}{\!:\!}

\newcommand{\toppos}{\epsilon} % top position

\def\res{\mathrel{\vert\grave{ }}}

\def \tuple#1{\langle #1 \rangle}

\long\def\comment#1{}

\newcommand{\none}{\bot}

\usepackage{xcolor}

\newcommand{\black}[1]{{\color{black} #1}}

\newcommand{\blue}[1]{{\color{blue} #1}}

\newcommand{\ptrs}{\mbox{PTRS}}
\newcommand{\pctrs}{\mbox{PCTRS}}
\newcommand{\ptr}{\cS}
\newcommand{\sel}{\kappa}
\newcommand{\prob}{\mathsf{p}}
\newcommand{\seld}{\mathbb{S}}

\newcommand{\pto}{\rightharpoondown}
\newcommand{\ppto}[1]{\stackrel{#1}{\rightharpoondown}}
\newcommand{\pptos}[1]{\stackrel{#1}{\pto}^{\raisebox{-.5em}{$\scriptstyle \ast$}}}

\newcommand{\dprob}{\cD^p}
\newcommand{\dreg}{\cD^r}

\usepackage{lineno}

\begin{document}

\title{A Distribution Semantics for \\Probabilistic Term Rewriting%
\thanks{This work has been partially supported by grant PID2019-104735RB-C41
funded by MICIU/AEI/ 10.13039/501100011033, by the 
\emph{Generalitat Valenciana} under grant CIPROM/2022/6 
(FassLow), and by TAILOR, a project funded by EU Horizon 2020 
research and innovation programme under GA No 952215.
}
}

\author{Germ\'an Vidal}

\institute{
VRAIN, Universitat Polit\`ecnica de Val\`encia, Spain\\
  \email{gvidal@dsic.upv.es}
}

\titlerunning{A Distribution Semantics for Probabilistic Term Rewriting}

\maketitle

\pagestyle{headings}

\begin{abstract}
Probabilistic programming is becoming increasingly popular 
thanks to its ability to specify problems with a certain 
degree of uncertainty. 
In this work, we focus on term rewriting, a 
well-known computational formalism. In particular, 
we consider systems that combine traditional rewriting 
rules with probabilities. Then, we define a novel
``distribution semantics'' for such systems that can be used
to model the probability of reducing a term to some value.
We also show how to compute a set of ``explanations'' for 
a given reduction, which can be used to compute its
probability in a more efficient way. 
Finally, we illustrate our approach with 
several examples and outline a couple of extensions that 
may prove useful to improve the expressive power of 
probabilistic rewrite systems.
\end{abstract}

%%%%%%%%%%%%%%%%%%%%%%%%%%%%%%%%%%%%%%%%%%%%%%%%%%%%%%%%%%%%%%%%%%%
\section{Introduction} \label{sec:intro}

Term rewriting \cite{Terese03} is a widely recognized 
computational formalism with many applications, ranging 
from providing theoretical foundations for programming
languages to problem modeling 
and developing analysis and 
verification techniques.
The addition of probabilities to a programming language
or formalism has proven useful for 
modeling domains with complex relationships and
uncertainty.
Surprisingly, however, we find only
a few proposals for \emph{probabilistic} term rewriting.
Among the various approaches to incorporating probabilities 
into term rewriting, the proposal by Bournez et al.\ 
\cite{BK02,BG05,BG06} is perhaps the most well-known. 
It introduces the concept PARS (Probabilistic Abstract
Reduction System) and its instance \ptrs\ (Probabilistic Term 
Rewrite System), together with the corresponding notions 
of probabilistic (abstract) reduction.
Intuitively speaking, in this line of work, 
a \emph{probabilistic term rewrite system} 
(\ptrs) is a collection of probabilistic 
rules of the form
\[
\blue{l \to p_1\cons r_1; \ldots;p_m\cons r_m
\hspace{2ex}
\mbox{\black{with}}~\sum_{i\in\{1,\ldots,m\}} p_i = 1
}
\]
where $l,r_1,\ldots,r_m$ are terms, $p_1,\ldots,p_m$ are
real numbers in $[0,1]$, and 
the right-hand side, $p_1\cons r_1, \ldots,p_m\cons r_m$,
represents a probability distribution over 
terms.\footnote{Actually, this is
called a \emph{multi-distribution} in \cite{ALY20} since
it is a multiset and repeated occurrences of the same term
are allowed.} 
This stream of work has mainly focused on the
termination properties of PTRSs, studying notions like
\emph{almost-sure termination} (AST), i.e., determine whether the
probability of termination is 1 \cite{BK02}, 
as well as several variants
of this notion (like Positive AST \cite{Sah78,BG05} 
or Strong AST \cite{ALY20}).
Besides the work of Bournez et al, confluence and 
termination of PTRSs has been considered, e.g., in 
\cite{ALY20,DCM18,KC17,KC19,KFG24,KDG24,Faggian22}.

In this paper, though, we consider a different topic within
probabilistic rewriting. In particular, we aim at defining
a novel \emph{distribution semantics} inspired to 
Sato's distribution semantics for probabilistic 
logic programs \cite{Sat95}.
Roughly speaking, Sato defines a distribution over the
\emph{least models} of a probabilistic 
logic program, thus generalizing
the traditional least model semantics. 
Here, we apply similar ideas in order
to define a distribution over the standard
TRSs that can be obtained from a PTRS by selecting a particular
choice of each probabilistic rewrite rule.
Consider, for instance, a simple PTRS that only includes the following probabilistic rules:
\[
\blue{\begin{array}{lrcllrcl}
R_1: & \mathsf{coin1} & \to & 0.5\cons \mathsf{heads}; 0.5\cons\mathsf{tails}\\
R_2: & \mathsf{coin2} & \to & 0.6\cons\mathsf{heads}; 0.4\cons \mathsf{tails}
\end{array}
}
\]
In this model, we have two coins: a fair coin, \textsf{coin1}, that 
lands on heads and on tails with equal probability, and a biased 
coin, \textsf{coin2}, that lands on heads with more 
probability than on tails.
Here, we could say that this PTRS encodes the 
following four ``worlds'' (represented by standard TRSs):
\[
\blue{\begin{array}{l@{~~~~~}l}
\cW_1 = \left\{
\begin{array}{rcl}
\mathsf{coin1} & \to & \mathsf{heads}\\
\mathsf{coin2} & \to & \mathsf{heads}\\
\end{array}\right\} 
&
\cW_2 = \left\{
\begin{array}{rcl}
\mathsf{coin1} & \to & \mathsf{heads}\\
\mathsf{coin2} & \to & \mathsf{tails}\\
\end{array}\right\} \\[4ex]
\cW_3 = \left\{
\begin{array}{rcl}
\mathsf{coin1} & \to & \mathsf{tails}\\
\mathsf{coin2} & \to & \mathsf{heads}\\
\end{array}\right\} 
&
\cW_4 = \left\{
\begin{array}{rcl}
\mathsf{coin1} & \to & \mathsf{tails}\\
\mathsf{coin2} & \to & \mathsf{tails}\\
\end{array}\right\} \\
\end{array}
}
\]
by considering all possible combinations of choices for the 
probabilistic rules.

The probability of each TRS is then 
determined by the product of the
probabilities of the choices made in each rule. 
Therefore, in this
example, our distribution would be as follows: $P(\cW_1) = 
0.5*0.6 = 0.3$, $P(\cW_2) = 0.5*0.4 = 0.2$, 
$P(\cW_3) = 0.5*0.6 = 0.3$, and $P(\cW_4) = 0.5*0.4 = 0.2$,
with $P(\cW_1) + P(\cW_2) + P(\cW_3) + P(\cW_4) = 1$.

Then, we will focus on calculating \emph{the 
probability that a term $s$ is reduced to another term 
$t$} (usually a \emph{value}), i.e., a typical 
reachability problem, which we denote by $s\to^\ast t$,
where $\to^\ast$ is the reflexive and transitive closure
of the rewrite relation $\to$ associated to some 
(probabilistic) rewrite system. For instance,
the probability that $\mathsf{coin1}$ reduces to 
$\mathsf{heads}$,
denoted by $P(\mathsf{coin1} \to^\ast \mathsf{heads})$,
is trivially 0.5 using the rule $R_1$ above.

In general, the probability that $s$ reduces to $t$ in
a PTRS is obtained from the sum of the probabilities
of the worlds where such a reduction can be proved.
E.g., $P(\mathsf{coin1} \to^\ast \mathsf{heads})
= P(\cW_1) + P(\cW_2) = 0.3 + 0.2 = 0.5$. 

A distinguishing feature of our proposal is that the 
application of a probabilistic rule is considered a 
\emph{one-time event}, in the sense that the selected 
choice cannot be changed later in the same reduction.  
In particular, the
tuple $\tuple{\mathsf{coin1},\mathsf{coin1}}$ can only
be reduced to either $\tuple{\mathsf{heads},\mathsf{heads}}$
or $\tuple{\mathsf{tails},\mathsf{tails}}$. 
For example, $P(\tuple{\mathsf{coin1},\mathsf{coin1}}
\to^\ast \tuple{\mathsf{heads},\mathsf{heads}})
%= P(\mathsf{coin1} \to^\ast \mathsf{heads})
= P(\cW_1) + P(\cW_2) = 0.3 + 0.2 = 0.5$,
while $P(\tuple{\mathsf{coin1},\mathsf{coin1}}
\to^\ast \tuple{\mathsf{heads},\mathsf{tails}}) = 0$
since there is no world where $\mathsf{coin1}$ 
can be reduced both to 
$\mathsf{heads}$ and to $\mathsf{tails}$.

Our approach  
contrasts to previous work (based on that of
Bournez et al.\ \cite{BK02,BG05,BG06})
where a different choice can be taken every time
a probabilistic rule is applied in a reduction. 
Moreover, in their setting, the
probability of a reduction is obtained from
the product of the probabilities of the steps, no
matter if the same probabilistic rule is 
applied several times.
Thus, the probability of the reduction
$\tuple{\mathsf{coin1},\mathsf{coin1}}
\stackrel{0.5}{\to}
\tuple{\mathsf{heads},\mathsf{coin1}} 
\stackrel{0.5}{\to} \tuple{\mathsf{heads},\mathsf{heads}}$
would be 0.5*0.5 = 0.25
in the approach Bournez et al., 
accounting for the two applications
of rule $R_1$. 
On the other hand, $\tuple{\mathsf{coin1},\mathsf{coin1}}$ 
could also be reduced to 
$\tuple{\mathsf{heads},\mathsf{tails}}$ with the same
probability.

We do not mean to say that one semantics is \emph{better} 
than the other. For some problems, one semantics could
be more appropriate, and vice versa. For example, 
as we will show in Example~\ref{ex:bayesian}, our proposal 
allows one to represent a Bayesian network in a natural way.
The main contributions of this work 
can be summarized as follows:
\begin{itemize}
\item First, we introduce a definition of a PTRS 
that allows us to define in a simple way a distribution 
semantics for probabilistic systems.

\item Next, we focus on computing the probability that 
a term $s$ reduces to a term $t$, 
in symbols $P(s\to^\ast t)$, and present a 
declarative definition for $P(s\to^\ast t)$. 
Unfortunately, this definition is not always feasible
in practice and, thus, we also propose an extension of 
term rewriting that allows us to generate a (complete) set of 
``explanations'' associated to a reduction $s\to^\ast t$. 
From these explanations, the probability 
$P(s\to^\ast t)$ can be computed 
using well-known techniques.

\item Finally, we illustrate the usefulness 
our proposal through several 
examples and introduce
a couple of extensions for PTRSs that can 
help to increase their expressive power.
\end{itemize}
The distribution semantics has been quite successful 
in the context of logic programming, where several 
languages have emerged that follow it, e.g.,
ProbLog \cite{dRKT07}, 
Logic Programs with Annotated Disjunctions \cite{VVB04}, 
Independent Choice Logic \cite{Poo97},
PRISM \cite{SK97}, etc.
Therefore, we think that it is worth considering a 
similar approach in the field of term rewriting too.

%%%%%%%%%%%%%%%%%%%%%%%%%%%%%%%%%%%%%%%%%%%%%%%%%%%%%%%%%%%%%%%%%%%%%%%%%%%%
\section{Standard Term Rewriting} \label{sec:trs}

To be self-contained, in this section we recall some 
basic concepts from term rewriting.
We refer the reader to, e.g., \cite{BN98} and \cite{Terese03},
for further details.

\subsection*{Terms and Substitutions}

A \emph{signature} $\cF$ is a finite 
set of ranked function symbols.  Given a set of
variables $\cV$ with $\cF\cap\cV=\emptyset$, we denote the domain of
\emph{terms} by $\cT(\cF,\cV)$.  We use $\mathsf{f},\mathsf{g},\ldots$
to denote functions and $x,y,\ldots$ to denote variables.
Positions are used to address the nodes of a term viewed as a tree. A
\emph{position} $\pi$ in a term $t$, in symbols $\pi\in\pos(t)$, is
represented by a finite sequence of natural numbers, where $\toppos$
denotes the root position.
We let $t|_\pi$ denote the \emph{subterm} of $t$ at position $\pi$ and
$t[s]_\pi$ the result of \emph{replacing the subterm} $t|_\pi$ by the term
$s$.
The set of variables appearing in term $t$ is
denoted by $\Var(t)$.  We also let
$\Var(t_1,\ldots,t_n)$ denote $\Var(t_1)\cup\cdots\cup\Var(t_n)$.  A
term $t$ is \emph{ground} if $\Var(t) = \emptyset$.

A \emph{substitution} $\sigma : \cV \mapsto \cT(\cF,\cV)$ is a mapping
from variables to terms such that $\dom(\sigma) = \{x \in \cV \mid x
\neq \sigma(x)\}$ is its domain.  
A substitution $\sigma$ is \emph{ground} if $x\sigma$ is ground for
all $x\in\dom(\sigma)$.
Substitutions are extended to morphisms from $\cT(\cF,\cV)$ to
$\cT(\cF,\cV)$ in the natural way.  We denote the application of a
substitution $\sigma$ to a term $t$ by $t\sigma$ rather than
$\sigma(t)$. The identity substitution is denoted by $\id$.
We let ``$\circ$'' denote the composition of substitutions, i.e.,
$(\sigma\circ\theta)(x) = (x\theta)\sigma = x\theta\sigma$.
The \emph{restriction} $\theta\!\res_V$ of a substitution $\theta$ to
a set of variables $V$ is defined as follows: $x\theta\!\res_{V} =
x\theta$ if $x\in V$ and $x\theta\!\res_V = x$ otherwise. 

\subsection*{Term Rewriting Systems}

A finite set of rewrite rules $l \to r$ such that $l$ is a nonvariable term
and $r$ is a term whose variables appear in $l$ is called a \emph{term
  rewriting system} (TRS for short); terms $l$ and $r$ are called the
left-hand side and the right-hand side of the rule, respectively. 
Given a TRS $\cR$
over a signature $\cF$, the \emph{defined} symbols $\cD_\cR$ are the
root symbols of the left-hand sides of the rules and the
\emph{constructors} are $\cC_\cR = \cF \;\backslash\; \cD_\cR$.
\emph{Constructor terms} of $\cR$ are terms over $\cC_\cR$ and $\cV$,
denoted by $\cT(\cC_\cR,\cV)$.  We sometimes omit $\cR$ from $\cD_\cR$
and $\cC_\cR$ if it is clear from the context.  A substitution
$\sigma$ is a \emph{constructor substitution} (of $\cR$) if $x\sigma
\in \cT(\cC_\cR,\cV)$ for all variables $x$.
 
For a TRS $\cR$, we define the associated rewrite relation $\to_\cR$
as the smallest binary relation on terms satisfying the following: given terms
$s,t\in\cT(\cF,\cV)$, we have $s \to_\cR t$ iff there exist a position
$\pi$ in $s$, a rewrite rule $l \to r \in \cR$, and a substitution
$\sigma$ such that $s|_\pi = l\sigma$ and $t = s[r\sigma]_\pi$; the rewrite
step is sometimes denoted by $s \to_{\pi,l\to r} t$ to make explicit the
position and rule used in this step.  
The instantiated left-hand side $l\sigma$ is called a \emph{redex}.
A term $s$ is called \emph{irreducible} or in \emph{normal form}
with respect to a TRS $\cR$ if there is no term $t$ with $s \to_\cR t$;  
we often omit ``with respect to $\cR$'' if it is clear from the context.  
A \emph{derivation} is a (possibly empty) sequence of rewrite steps.
Given a binary relation $\to$, we denote by $\rightarrow^{\ast}$ its
reflexive and transitive closure, i.e., $s \to^\ast_\cR t$ means that
$s$ can be reduced to $t$ in $\cR$ in zero or more steps;
furthermore, we let $s\to^! t$ if $s\to^\ast t$, and, moreover,
$t$ is a normal form. 

%%%%%%%%%%%%%%%%%%%%%%%%%%%%%%%%%%%%%%%%%%%%%%%%%%%%%%%%%%%%%%%%%%%
\section{A Distribution Semantics for Probabilistic TRSs} \label{sec:ptrs}

In this section, we first present an extension of rewrite
systems to model uncertainty. Then, we define a 
\emph{distribution semantics} for probabilistic systems.
Finally, we
consider the problem of computing the probability
of $s \to^\ast t$ for some given terms $s,t$.

\subsection{Probabilistic TRSs}

In the following, we assume that the domain of defined
function symbols
$\cD$ is partitioned into a set $\dprob$ of 
\emph{probabilistic} functions and a set 
$\dreg$ of \emph{regular} (non-probabilistic) functions, 
which are disjoint, i.e., $\cD=\dprob\uplus\dreg$.
A \emph{probabilistic rule} has the form
\[
\blue{R: l \to p_1\cons r_1;\ldots;p_n\cons r_n}
\]
where $R$ is a label that uniquely identifies the 
rule,\footnote{We will omit labels when they are not relevant.}
$l,r_1,\ldots,r_n$ are \emph{ground} terms
and $p_1,\ldots,p_n$
are real numbers in the interval $[0, 1]$ (their respective probabilities)
such that $\sum_{i=1}^n p_i \sleq 1$. 
When $\sum_{i=1}^n p_i < 1$, we 
implicitly assume that a special term $\none$ 
is added to the right-hand side
of the rule, where $\none$ 
is a fresh constructor symbol of arity zero (a constant)
which does not occur in the
original system, with associated probability 
$1 - \sum_{i=1}^n p_i$. 
Here, choosing $\bot$ is equivalent to not 
reducing the selected term (nor any of its descendants).
Thus, in the following, we
assume w.l.o.g.\ that $\sum_{i=1}^n p_i = 1$ for
all probabilistic rules.
A \emph{probabilistic term rewrite 
system}\footnote{As long as there is no confusion, 
we will use the same terminology (\ptrs) for probabilistic 
term rewrite systems as in the works based on the 
approach of \cite{BK02,BG05,BG06} despite the differences.} 
(\ptrs) $\ptr = \ptr_p\uplus\ptr_r$ with 
$\cD_\ptr = \dprob_\ptr\uplus\dreg_\ptr$ 
consists of a set of probabilistic rules $\ptr_p$ 
defining the defined function symbols in $\dprob_\ptr$
and a set of regular rewrite rules $\ptr_r$
defining the defined function symbols in 
$\dreg_\ptr$.

\begin{example}  \label{ex:coins}
Consider the following \ptrs\ with two probabilistic
rules and two regular rules:
\[\hspace{-2ex}
\blue{\begin{array}{lrcl@{~~~}ll}
R_1: & \mathsf{coin1} & \to & 0.5\cons\mathsf{heads}; 0.5\cons\mathsf{tails} &
R_3: & \mathsf{main}  \to  \mathsf{flip2coins}(\mathsf{coin1},\mathsf{coin2}) \\
R_2: & \mathsf{coin2} & \to & 0.6\cons\mathsf{heads}; 0.4\cons\mathsf{tails} &
R_4: & \mathsf{flip2coins}(x,y)  \to  \mathsf{t2}(x,y)\\
%[1ex]
%\mathsf{twoCoins} & \to & \tuple{\mathsf{coin1},\mathsf{coin2}}\\
%\mathsf{switch}(\tuple{x,y}) & \to & \tuple{x,y}  \\ 
%\mathsf{switch}(\tuple{x,y}) & \to & \tuple{y,x}\\ 
\end{array}
}
\]
In this example, function $\mathsf{main}$ represents the
flipping of two coins. Functions $\mathsf{coin1}$
and $\mathsf{coin2}$ represent the two coins already
mentioned in the Introduction.
We have also a rewrite rule for the regular function
$\mathsf{flip2coins}$ that simply reduces to a tuple
$\mathsf{t2}(\_,\_)$
with the outcome of flipping the two coins.
\end{example}
Although the ground requirement for probabilistic 
rules may seem too restrictive, it is quite natural for
many practical problems. This condition is essential 
for the number of possible ``worlds'' to be finite, 
as we will see below. Nevertheless, in the next section 
we will consider several ways to get around this restriction.

\subsection{A Distribution Semantics for \ptrs s}

Let us now consider the definition of a distribution
semantics for \ptrs s. Essentially, each probabilistic rule 
$R : (l \to p_1\cons r_1;\ldots;p_n\cons r_n)$
represents a \emph{choice} between $n$
regular rules: $l \to r_1$,
\ldots, $l \to r_n$.
In other words, each probabilistic rule represents a 
\emph{random variable} of the model represented by the PTRS.
A particular choice is denoted 
by a tuple $(R,i)$, $i\in\{1,\ldots,n\}$, 
which is called an \emph{atomic choice}, and has as
associated probability $\prob(R,i)$, i.e., $p_i$
in the rule above.
Following \cite{Poo00}, we introduce the following notions:

\begin{definition}[composite choice, selection]
\label{def:compositechoice}
We say that a set $\kappa$ of atomic choices is \emph{consistent}
if it does not contain two atomic choices for the same 
probabilistic rule, i.e., it cannot contain 
$(R,i)$ and $(R,j)$ with $i\neq j$.

A set of consistent atomic choices is called a 
\emph{composite choice}. 

Two composite choices, $\kappa_1,\kappa_2$, are
\emph{incompatible} if $\kappa_1\cup\kappa_2$ is 
not consistent, i.e., we have $(R,i)\in\kappa_1$ and
$(R,j)\in\kappa_2$ with $i\neq j$.

A set $K$ of composite choices is \emph{mutually incompatible} 
if for all $\kappa_1\in K$, $\kappa_2\in K$, 
$\kappa_1\neq\kappa_2$ implies $\kappa_1\cup\kappa_2$ 
is not consistent.

A composite choice is called a \emph{selection}
when it includes an atomic choice for
each probabilistic rule of the \ptrs.
We let $\seld_\ptr$ denote the set of all 
possible selections for a given \ptrs\ $\ptr$.
\end{definition}
Each selection $\sel$ identifies a (possible) \emph{world}
which contains a regular rule $l \to r_i$ for each 
atomic choice $(R,i)\in \sel$, together with the rules
for regular functions. Formally:

\begin{definition}[world]
  Let $\ptr$ be a \ptrs. A
  selection $\sel\in\seld_\ptr$ 
  identifies a \emph{world} $\cW_\sel$ defined as follows:
  \[
  \blue{\cW_\sel
  = \{ l \to r_i \mid
  R:(l\to p_1\cons r_1;\ldots;p_n\cons r_n)\in\ptr_p\wedge
  (R,i)\in \sel\} \cup \ptr_r}
  \]
  The probability of world $\cW_\sel$ 
  is then given by 
  \[
  \blue{
  P(\cW_\sel) =  P(\sel) = \prod_{(R,i)\in \sel} \prob(R,i)
  }
  \]
  We let $\mathbb{W}_\ptr$ 
  denote the 
  set of all possible worlds, 
  i.e., $\mathbb{W}_\ptr = \{\cW_\sel \mid 
  \sel\in\mathbb{S}_\ptr\}$.
  Here, $P(\cW)$ defines a probability distribution over 
  $\mathbb{W}_\ptr$. 
  By definition, the sum of the probabilities of 
  all possible worlds is equal to $1$.
\end{definition}
In this context, given a \ptrs\ $\ptr$,
we are interested in computing the probability 
of whether a term $s$ can be reduced to term $t$ in 
$\ptr$---a typical reachability problem---,
in symbols $P(s\to^\ast_\ptr t)$  or simply
$P(s\to^\ast t)$ when the \ptrs\ $\ptr$ is clear from the context.
This probability can be obtained by marginalization from
the joint distribution $P(s\to^\ast_\ptr t, \cW)$ as follows:

\begin{definition}[probability of $s\to^\ast t$]
Let $\ptr$ be a \ptrs\ and $s,t$ be terms. Then, the
probability of $s\to^\ast_\ptr t$ is given by
\[
\blue{P(s\to^\ast_\ptr t) = \sum_{\cW\in \mathbb{W}_\ptr} 
P(s\to^\ast_\ptr t,\cW) = 
\sum_{\cW\in \mathbb{W}_\ptr} P(s\to^\ast_\ptr t\mid \cW) \cdot 
P(\cW)}
\]
where $P(s\to^\ast_\ptr t\mid \cW) = 1$ if there exists a 
derivation $s\to^\ast_\cW t$  
and $P(s\to^\ast_\ptr t \mid \cW) = 0$  otherwise.
\end{definition}
Intuitively speaking, the probability of $s\to^\ast t$ is 
equal to the sum of the probabilities of all the worlds 
where the term $s$ can be reduced to the term $t$.
Often, we will be interested in knowing the probability 
$P(s \to^\ast t)$ where $t$ is a normal form (a possible 
``value'' of $s$).

\begin{example} \label{ex:coins2}
Consider again the \ptrs\ of Example~\ref{ex:coins}.
Here, we have four possible worlds:
\[
\blue{
\begin{array}{l@{}l@{}l@{}l@{}l@{}l}
\cW_1 & = & \{ \mathsf{coin1} \to \mathsf{heads}, & \mathsf{coin2}\to\mathsf{heads}
%, & \mathsf{twoCoins} \to \tuple{\mathsf{coin1},\mathsf{coin2}} 
&  \}\\
\cW_2 & = & \{ \mathsf{coin1} \to \mathsf{heads}, & \mathsf{coin2}\to\mathsf{tails}
%, & \mathsf{twoCoins} \to \tuple{\mathsf{coin1},\mathsf{coin2}} 
& \}\\
\cW_3 & = & \{ \mathsf{coin1} \to \mathsf{tails}, & \mathsf{coin2}\to\mathsf{heads}
%, & \mathsf{twoCoins} \to \tuple{\mathsf{coin1},\mathsf{coin2}} 
& \}\\
\cW_4 & = & \{ \mathsf{coin1} \to \mathsf{tails}, & \mathsf{coin2}\to\mathsf{tails}
%,& \mathsf{twoCoins} \to \tuple{\mathsf{coin1},\mathsf{coin2}} 
& \}\\
\end{array}
~\bigcup~\left\{\begin{array}{l}
\mathsf{main}\to\mathsf{flip2coins}(\mathsf{coin1},\mathsf{coin2})\\
\mathsf{flip2coins}(x,y)  \to  \mathsf{t2}(x,y)\\
%\mathsf{switch}(\tuple{x,y}) & \to & \tuple{x,y} \\
%\mathsf{switch}(\tuple{x,y}) & \to & \tuple{y,x} \\
\end{array}
\right\}
}
\]
with associated probabilities $P(\cW_1) = 0.5*0.6 = 0.3$,
$P(\cW_2) = 0.5*0.4 = 0.2$, $P(\cW_3) = 0.5*0.6 = 0.3$,
and $P(\cW_4) = 0.5*0.4 = 0.2$.
Then, for instance, the probability 
of reducing 
$\mathsf{main}$
to $\mathsf{t2}(\mathsf{heads},\mathsf{heads})$ is given by
$P(\mathsf{main}\to^\ast \mathsf{t2}(\mathsf{heads},\mathsf{heads}))
= P(\cW_1) = 0.3$.
Let us now add the following (nondeterministic) regular 
rules to the considered \ptrs:
\[
\blue{\begin{array}{rcl}
R_5:~~\mathsf{switch}(\mathsf{t2}(x,y)) & \to & \mathsf{t2}(x,y) \\
R_6:~~\mathsf{switch}(\mathsf{t2}(x,y)) & \to & \mathsf{t2}(y,x) \\
\end{array}}
\]
The number of worlds and their probabilities do not change
since the probabilistic rules are the same; the only
difference is that, now, rules $R_5$ and $R_6$ are added 
to every world.
Consider now the probability of reducing 
$\mathsf{switch}(\mathsf{main})$
to $\mathsf{t2}(\mathsf{heads},\mathsf{tails})$, i.e.,
the probability of getting one heads and one tails, 
no matter which coin. 
Here, we have the following derivation $D_1$ in world $\cW_2$:
\[
\blue{
\begin{array}{lll}
\mathsf{switch}(\mathsf{main}) & \to & 
\mathsf{switch}(\mathsf{flip2coins}(\mathsf{coin1},\mathsf{coin2})) \\
& \to & \mathsf{switch}(\mathsf{t2}(\mathsf{coin1},\mathsf{coin2}))\\
& \to & \mathsf{t2}(\mathsf{coin1},\mathsf{coin2})\\
& \to & \mathsf{t2}(\mathsf{heads},\mathsf{coin2})\\
& \to & \mathsf{t2}(\mathsf{heads},\mathsf{tails})
\end{array}
}
\]
and the following one, $D_2$, in world $\cW_3$:
\[
\blue{
\begin{array}{lll}
\mathsf{switch}(\mathsf{main}) & \to &
\mathsf{switch}(\mathsf{flip2coins}(\mathsf{coin1},\mathsf{coin2}))\\ 
& \to & \mathsf{switch}(\mathsf{t2}(\mathsf{coin1},\mathsf{coin2}))\\
& \to & \mathsf{t2}(\mathsf{coin2},\mathsf{coin1})\\
& \to & \mathsf{t2}(\mathsf{heads},\mathsf{coin1})\\
& \to & \mathsf{t2}(\mathsf{heads},\mathsf{tails})
\end{array}
}
\]
Thus, 
$P(\mathsf{switch}(\mathsf{main}) \to^\ast \mathsf{t2}(\mathsf{heads},\mathsf{tails})) = P(\cW_2) + P(\cW_3) = 0.2 + 0.3 = 0.5$,
i.e., the probability of getting heads and tails with the 
two coins does not depend on whether one of them is biased
(it is always 50\%).
\end{example}
We note that $P(s\to^\ast t)$ does not depend on the number of
possible derivations from $s$ to $t$ in each world, but only
on the number of worlds where the derivation is possible.

The usual properties over standard TRSs can be lifted to
\ptrs\ straightforwardly by considering the associated
worlds. E.g., we say that a \ptrs\ $\ptr$
is terminating (resp.\ confluent) if every world 
$\cW\in\mathbb{W}_\ptr$
is terminating (resp.\ confluent). 
In general:

\begin{definition} \label{def:lifting}
  Let $\ptr$ be a \ptrs.
  We say that a property $P$ holds for \ptrs\ $\ptr$
  if $P$ holds for each world $\cW\in\mathbb{W}_\ptr$.
\end{definition}
Unfortunately, the number of worlds grows exponentially 
with the number of random variables represented in the 
model, i.e., with the number of (ground instances of the) 
probabilistic rules in the \ptrs. For example, $n$ ground
rules with only two choices give rise to $2^n$ worlds. 
This makes computing $P(s \to^\ast t)$ by summing up
the probabilities of the worlds where $s$ can be reduced 
to $t$ intractable. 

\subsection{Knowledge Compilation for Probabilistic Inference}

In the following, we present an alternative method based
on so-called \emph{knowledge compitation} \cite{DM02}.
Essentially, this approach is based on compiling a Boolean
formula (representing a set of worlds) 
into a representation where inference is tractable. 
This approach has been used, e.g., for
performing inference in ProbLog \cite{dRKT07}, where the 
process consists basically in the following two steps:
\begin{enumerate}
 \item First, a propositional formula that represents 
 the worlds where a given query is true is computed.
 For this purpose, a \emph{grounding} stage is usually
 required. Then, some variant of SLD
 resolution is used to evaluate the query and produce
 the so-called \emph{explanations} (see below), 
 a representation of the
 worlds where the query is true.

 \item Then, the propositional formula is compiled into
 a binary decision diagram (BDD) \cite{Bry86} or some 
 other similar data structure where 
 \emph{weighted model counting} \cite{CD08} 
 can be done efficiently
 (often, in time linear in the size of the compiled 
 representation). In this context, 
 weights are associated to the probabilities 
 of the original program. 
\end{enumerate}
Obviously, the complexity of the inference
is the same (namely, the problem is PP-complete 
\cite{CM17}) but now it is moved into the compilation 
stage. Moreover, it has been experimentally shown 
that the approach
works well in many practical cases, scaling up to 
queries with a few hundred thousand explanations
\cite{VKDMR16}.
Alternatively, when exact inference is too expensive, 
we can still resort to \emph{approximate} inference
(some techniques for approximate inference are briefly 
described at the end of Section~\ref{sec:computing}).

In the remainder of this section, 
we present a knowledge compilation
approach to probabilistic inference in our setting.
Similarly to \cite{Poo93,Poo00} in the context of 
a formalism based on Horn clauses, we consider 
the notion of \emph{explanation}.
First, we say that a selection $\sel$ 
\emph{extends} a composite choice $\kappa'$ 
if $\sel\supseteq\kappa'$; analogously, a world $\cW$
is \emph{compatible} with a composite choice 
$\kappa'$ if there is a selection $\sel$ that extends 
$\kappa'$ and $\cW=\cW_\sel$.
In general, a composite choice $\kappa'$ identifies 
the set of worlds $\cW_{\kappa'}$ that are compatible with $\kappa'$,
i.e., 
$\cW_{\kappa'} = \{ \cW_\sel \mid \sel\in\mathbb{S}_\ptr
\wedge \sel\supseteq \kappa' \}$.
Moreover, given a (finite) set of composite choices $K$, we let
$\cW_K = \cup_{\kappa\in K}~ \cW_\kappa$.

\begin{definition}[explanation, covering]
  Let $\ptr$ be a \ptrs\ and $\kappa$ a composite choice
  for the probabilistic rules of $\ptr$. Given terms
  $s,t$, we say that $\kappa$ is an \emph{explanation}
  for $s\to^\ast_\ptr t$ iff there exists a derivation
  $s\to^\ast t$ in each world that is compatible with $\kappa$,
  i.e., there is a derivation
  $s\to^\ast_\cW t$ for all $\cW \in \cW_\kappa$.

  Let $K$ be a (finite) set of explanations for $s\to^\ast_\ptr t$.
  We say that $K$ is \emph{covering} for
  $s\to^\ast_\ptr t$ if for all $\cW\in \mathbb{W}_\ptr$ 
  such that $s \to^\ast_{\cW} t$ we have $\cW\in\cW_K$.
\end{definition}
A set of covering explanations for $s\to^\ast_\ptr t$ can be
seen as a compact representation of all the worlds where
$s$ can be reduced to $t$. In some cases, using this 
set of explanations---rather than all 
possible worlds---can make a 
significant difference regarding the time needed to
compute the probability of $s\to^\ast_\ptr t$.

Now, in order to compute a 
covering set of explanations
for a given reachability problem, 
we introduce an extension of rewriting
over pairs $\tuple{t,\kappa}$, where $t$ is a term and
$\kappa$ is a composite choice. 

\begin{definition}[probabilistic rewriting]
Let $s,t$ be terms and $\kappa,\kappa'$ be composite choices.
A \ptrs\ $\ptr$ induces a \emph{probabilistic rewrite} 
relation $\pto_\ptr$ 
where $\tuple{s,\kappa} \pto_\ptr \tuple{t,\kappa'}$
if either
\begin{itemize}
\item there is a position
$\pi$ in $s$, a regular rewrite rule $l \to r \in \ptr$, and a 
substitution $\sigma$ such that $s|_\pi = l\sigma$, 
$t = s[r\sigma]_\pi$, and $\kappa'=\kappa$, or

\item there exists a position
$\pi$ in $s$, a probabilistic rewrite rule 
$R : (l \to p_1\cons r_1;\ldots;p_n\cons r_n) \in \ptr$, and a 
substitution $\sigma$ such that $s|_\pi = l\sigma$, 
$t = s[r_i\sigma]_\pi$, $i\in\{1,\ldots,n\}$,
and $\kappa' = \kappa\cup\{(R,i)\}$ is consistent.
\end{itemize}
We often label a probabilistic rewrite step with the 
probability of the choice made:
given 
$\tuple{s,\kappa} \stackrel{p}{\pto} 
\tuple{t,\kappa'}$,
we have $p=\prob(R,i)$ if 
$\kappa'\setminus\kappa = \{(R,i)\}$,
and $p=1$ otherwise.
Probabilities can then be lifted to derivations by
computing the product of the probabilities of their steps,
i.e., 
$\tuple{s_0,\kappa_0} \pptos{p} \tuple{s_n,\kappa_n}$ 
if $\tuple{s_0,\kappa_0} \ppto{p_1} \tuple{s_1,\kappa_1} 
\ppto{p_2} \ldots \ppto{p_n} \tuple{s_n,\kappa_n}$ and 
$p=p_1*p_2*\ldots*p_n$.
\end{definition}
It is interesting to note that we only label a step 
with probability $\prob(R,i)$ when the rule $R$ is 
probabilistic and, moreover, it is the \emph{first time} 
it is applied in the derivation. 
Therefore, if we apply the same probabilistic rule 
several times in a derivation, we only count the 
associated probability once. 
Furthermore, the consistency requirement avoids
applying the same rule with a different choice.

The next result proves that our definition of probabilistic
rewriting is indeed sound and complete regarding the
computation of explanations.

\begin{theorem} \label{th:correctness}
  Let $\ptr$ be a \ptrs\ and let $s,t$ be terms. Then,
  for all selections $\sel\in\mathbb{S}_\ptr$, we have  
  $s \to^\ast_{\cW_\sel} t$
  iff
  $\tuple{s,\emptyset} \pto^\ast_\ptr \tuple{t,\kappa'}$
  with $\sel\supseteq\kappa'$.
\end{theorem}

\begin{proof}
First, we consider the ``only if'' direction. Let $s \to^\ast_{\cW_\sel} t$. Since $\sel$ is
consistent by definition, we can construct a derivation
$\tuple{s,\emptyset} \pto^\ast_\ptr \tuple{t,\kappa'}$ 
that mimicks the same steps of $s \to^\ast_{\cW_\sel} t$.
Trivially, $\kappa'$ is then a (consistent) composite choice
with $\sel\supseteq\kappa'$, as required.

Consider now the ``if'' direction. Let 
$\tuple{s,\emptyset} \pto^\ast_\ptr \tuple{t,\kappa'}$
be a derivation with the probabilistic rewrite relation.
By definition, $\kappa'$ is a (consistent) composite choice
and, thus, $s \to^\ast_{\cW} t$ can be proved in every
world that is compatible with $\kappa'$, since the 
relevant choices (those for the probabilistic rules 
used in the derivation) are the same. \qed
\end{proof}
In the following, we let $\mathbb{E}(s\to^\ast_\ptr t)$
denote the set of explanations computed by probabilistic
rewriting. Formally:

\begin{definition} \label{def:explanations}
  Let $\ptr$ be a \ptrs\ and $s,t$ be terms. Then, we define
  the explanations of $s\to^\ast t$ as follows:
  \[
\blue{
\mathbb{E}(s\to^\ast_\ptr t) = \{ \kappa \mid \tuple{s,\emptyset} \pto^\ast_\ptr \tuple{t,\kappa}\}
}
\]
Furthermore, the probability of an explanation is equal
to the probability of the associated derivation, i.e.,
given 
$D = \tuple{s,\emptyset} \pto^\ast_\ptr \tuple{t,\kappa}$, 
we have
\[
\blue{
P(\kappa) = P(D) = \prod_{(R,i)\in \kappa} \prob(R,i).
}
\]
\end{definition}
As a consequence of Theorem~\ref{th:correctness},
we have the following result:

\begin{corollary} \label{cor:covering}
  Let $\ptr$ be a \ptrs\ and let $s,t$ be terms. Then,
  $\mathbb{E}(s\to^\ast_\ptr t)$ is a covering set of explanations
  for $s\to^\ast_\ptr t$.
\end{corollary}

\begin{example} \label{ex:sumprob}
 Consider again the \ptrs\ of Example~\ref{ex:coins2} and the
 two derivations $D_1$ and $D_2$ from 
 $\mathsf{switch}(\mathsf{main})$ 
 to $\mathsf{t2}(\mathsf{heads},\mathsf{tails})$. 
 The corresponding
 derivations with the probabilistic rewrite relation, 
 $D'_1$ and $D'_2$, are shown in Figure~\ref{fig:sumprob}.
\begin{figure}[t]
 \[
\blue{
\begin{array}{lll}
\tuple{\mathsf{switch}(\mathsf{main}),\emptyset} & \pto &
\tuple{\mathsf{switch}(\mathsf{flip2coins}(\mathsf{coin1},\mathsf{coin2})),\emptyset} \\
& \pto & \tuple{\mathsf{switch}(\mathsf{t2}(\mathsf{coin1},\mathsf{coin2})),\emptyset}\\
& \pto & \tuple{\mathsf{t2}(\mathsf{coin1},\mathsf{coin2}),\emptyset}\\
& \stackrel{0.5}{\pto} & \tuple{\mathsf{t2}(\mathsf{heads},\mathsf{coin2}),\{(R_1,1)\}}\\
& \stackrel{0.4}{\pto} & \tuple{\mathsf{t2}(\mathsf{heads},\mathsf{tails}),\{(R_1,1),(R_2,2)\}}\\[2ex]

\tuple{\mathsf{switch}(\mathsf{main}),\emptyset} & \pto &
\tuple{\mathsf{switch}(\mathsf{flip2coins}(\mathsf{coin1},\mathsf{coin2})),\emptyset} \\
& \pto & \tuple{\mathsf{switch}(\mathsf{t2}(\mathsf{coin1},\mathsf{coin2})),\emptyset}\\
& \pto & \tuple{\mathsf{t2}(\mathsf{coin2},\mathsf{coin1}),\emptyset}\\
& \stackrel{0.6}{\pto} & \tuple{\mathsf{t2}(\mathsf{heads},\mathsf{coin1}),\{(R_2,1)\}}\\
& \stackrel{0.5}{\pto} & \tuple{\mathsf{t2}(\mathsf{heads},\mathsf{tails}),\{(R_1,2),(R_2,1)\}}
\end{array}
}
\]
\caption{Probabilistic derivations for 
$\mathsf{switch}(\mathsf{main}) \to^\ast\mathsf{t2}(\mathsf{heads},\mathsf{tails})$} \label{fig:sumprob}
\end{figure}
\end{example}
Computing a covering set of explanations
is generally undecidable. 
However, there are classes of \ptrs s 
for which this task becomes decidable, e.g., 
for terminating \ptrs s (i.e., for \ptrs s whose 
associated worlds are all terminating).
On the other hand, we can easily reuse existing results
from \emph{reachability analysis} in term rewriting
(see, e.g., \cite{FGT04,SY19}, and references therein) in order
to prune some derivations during
the computation of a covering set of explanations.
In particular, we are concerned with the 
\emph{infeasibility} of a reachability
problem: given a TRS $\cR$ and (ground) terms $s,t$,
determine if $s\to^\ast_\cR t$ is not provable.

In principle, we could trivially extend the notion of
reachability to \ptrs s by considering all the
associated worlds: a
reachability problem $s \to^\ast_\ptr t$ is unfeasible
in a \ptrs\ $\ptr$ if 
$s \to^\ast_\cW t$ is unfeasible
in all worlds $\cW\in\mathbb{W}_\ptr$.
Moreover, we can take advantage
of the \emph{partial} explanation computed so far 
in order to discard some possible worlds. 
In the following, given a \ptrs\ $\ptr$ and a
composite choice $\kappa$, we let 
$\ptr\!\res\!\kappa$
denote the \emph{restriction of $\ptr$ by $\kappa$}, 
which is defined as follows: 
$
\blue{
\ptr\!\res\!\kappa = \ptr_p^\kappa
%\cup (\ptr_p \setminus \ptr_p^\kappa) 
\cup \ptr_r
}
$,
where 
\[
\blue{
\begin{array}{lll}
\ptr_p^\kappa 
& = & \{ l \to r_i \mid
R:(l\to p_1\cons r_1;\ldots;p_n\cons r_n)\in\ptr_p\wedge
(R,i)\in \kappa\} \\
& & \cup\:  
\{ R:(l\to p_1\cons r_1;\ldots;p_n\cons r_n) \mid
R:(l\to p_1\cons r_1;\ldots;p_n\cons r_n)\in\ptr_p \\
&& \hspace{31ex}\wedge
\not\exists j~\mbox{such that}~(R,j)\in\kappa\}
\end{array}
}
\]
Intuitively speaking, $\ptr\!\res\!\kappa$ is equal to
$\ptr$ except for the probabilistic rules whose choice
already occurs in the partial explanation $\kappa$; these
probabilistic rules are replaced by regular rules by
removing all choices in their right-hand sides but the
selected one (and also removing its probability).

\begin{definition}[unfeasibility w.r.t.\ $\kappa$]
  Let $\ptr$ be a \ptrs\ and $s,t$ be terms.
  Given a composite choice $\kappa$, we say that
  the reachability problem $s \to^\ast_\ptr t$
  is unfeasible w.r.t.\ $\kappa$ if
  $s \to^\ast_{\ptr\res\kappa} t$ is infeasible.
\end{definition}
Therefore, during the computation of 
$\mathbb{E}(s\to^\ast_\ptr t)$, we can prune 
those derivations 
$\tuple{s,\emptyset} \pto^\ast_\ptr \tuple{s',\kappa}$ 
in which $s'\to^\ast_\ptr t$ is infeasible w.r.t.\ $\kappa$.

Reachability (and, thus, the infeasiblity of a reachablity
problem) is decidable for ground rewrite systems
\cite{Oya86}. 
There are also a number of techniques for proving the 
infeasibility of some reachability problems when 
certain conditions are met. Many of these techniques 
are based on some abstraction of the rewrite rules, 
together with a graph that represents
an approximation of the rewrite relation.
One of the earliest such techniques can be found in 
\cite{AFRV93}, in which the TRS abstraction consists 
of replacing each rule $l \to r$ by the set of rules 
$l \to \widehat{r_1}$,\ldots, $l\to \widehat{r_n}$, 
where  $r_1,\ldots,r_n$ are the subterms of $r$
rooted by a defined function symbol 
and $\widehat{f(t_1,\ldots,t_n)}
= f(\lfloor t_1\rfloor,\ldots,\lfloor t_n\rfloor)$
with $\lfloor t \rfloor = c(\lfloor t_1\rfloor,\ldots,\lfloor t_n\rfloor)$ if $c$ is a constructor symbol and 
$\lfloor t \rfloor = x$ otherwise, with $x$ a fresh variable.
Then, a \emph{graph of functional dependencies} is built by
\emph{connecting} $l$ to $\widehat{r_i}$, $i=1,\ldots,n$,
for each rule and, then, adding edges from $\widehat{r_i}$
to each left-hand side with which it unifies.
This graph of dependencies was used 
to prune infeasible equations in the context of 
basic narrowing \cite{Hul80} (an extension of term 
rewriting).

A similar technique was presented and popularized 
some years later in the context of the well-known
\emph{dependency pairs} method for proving the 
termination of rewrite systems \cite{Arts96,AG00}.
In this case, the elements of the 
graph are \emph{dependency pairs} of the form
$l^\sharp \to t^\sharp$ for each rule $l\to r$, where
$t$ is a subterm of $r$ rooted by a defined function
symbol and $t^\sharp = f^\sharp(t_1,\ldots,t_n)$
if $t=f(t_1,\ldots,t_n)$ and $f^\sharp$ is a fresh
function symbol. Then, dependency pairs 
$l^\sharp \to s^\sharp$ and 
$t^\sharp \to r^\sharp$ are \emph{connected} if
there is a substitution $\sigma$ such that
$s^\sharp\sigma \to^\ast t^\sharp\sigma$. 
This relation is often approximated, so that there
is an edge from $l^\sharp \to s^\sharp$ to 
$t^\sharp \to r^\sharp$ in the graph if
$\widehat{s^\sharp}$ unifies 
with $t^\sharp$, thus producing a similar result
as the graph from \cite{AFRV93} mentioned above.

The development of probabilistic rewrite strategies 
that prune derivations when the reachability problem is 
guaranteed to be infeasible 
is an interesting problem
that is outside the scope of this paper and 
is left as future work.

\subsection{Computing the Probability from a Covering Set of Explanations} \label{sec:computing}

Finally, in this section we focus on how to calculate the 
probability of a reachability problem from a set of 
covering explanations.
The following result is an easy consequence:

\begin{lemma} \label{lemma:sum}
  Let $\ptr$ be a \ptrs\ and $\kappa$ be a composite choice. 
  Then, $P(\kappa) = \sum_{\cW\in\cW_\kappa} P(\cW)$.
\end{lemma}

\begin{proof}
  We prove the claim by induction on the number $n$ of 
  probabilistic rules which do not occur in $\kappa$.
  If $n=0$ then $\kappa$ is a selection and the claim 
  follows. Consider $n>0$. Let $\cW'_1,\ldots,\cW'_m$
  be the worlds that are compatible with $\kappa$ 
  but not including the choices for some rule $R$.
  Assume now that the possible choices for rule $R$
  are $(R,1),\ldots,(R,u)$. By definition, we have
  $\prob(R,1)+\ldots+\prob(R,u) = 1$. Hence, 
  $\sum_{\cW\in\cW_\kappa} P(\cW)
  = \sum_{i=1,\ldots,m} P(\cW')*\sum_{j=1,\ldots,u} P(R,j)
  = \sum_{i=1,\ldots,m} P(\cW')*1 = P(\kappa)$ follows
  by the inductive hypothesis. \qed
\end{proof}
In the following, we let
$P(K)$ denote the sum of the probabilities of all the
worlds the are compatible with the explanations in $K$,
i.e., $P(K) = \sum_{\cW\in\cW_K} P(\cW)$.
Interestingly, for sets of mutually incompatible 
explanations (cf.\ Definition~\ref{def:compositechoice}), 
the following property holds:

\begin{lemma} \label{lemma:incom}
  Let $\ptr$ be a \ptrs\ and $K$ be a mutually
  incompatible set of explanations.
  Then, $P(K) = \sum_{k\in K} P(\kappa)$.
\end{lemma}

\begin{proof}
  The result is an straightforward consequence of
  \cite[Lemma 4.4]{Poo00} which states that, 
  if $K$ is mutually incompatible, then 
  there is no world in which more than one element 
  of $K$ is true. 
  Let $K = \{\kappa_1,\ldots,\kappa_n\}$.  
  Then, we have  
  $\cW_K = \cW_{\kappa_1} \cup \ldots \cup \cW_{\kappa_n}$
  with $\cW_{\kappa_1} \cap \ldots \cap \cW_{\kappa_n} 
  = \emptyset$. Hence, by Lemma~\ref{lemma:sum},
  $P(K) = 
  \sum_{\cW\in\cW_\kappa} P(\cW)
  = \sum_{\cW\in\cW_{\kappa_1}} P(\cW)
  + \ldots + \sum_{\cW\in\cW_{\kappa_n}} P(\cW)
  = P(\kappa_1) + \ldots + P(\kappa_n)$, which
  proves the claim. \qed
\end{proof}
Finally, when a set of explanations is both
covering and mutually incompatible, the probability
of a reachability problem can be computed in a
straightforward way:\footnote{A similar result for
ICL can be found in \cite{Poo00}.}

\begin{proposition} 
  Let $\ptr$ be a \ptrs\ and $s,t$ be terms.
  Let $K = \mathbb{E}(s\to^\ast_\ptr t)$.
  If $K$ is mutually incompatible, then
  $P(s\to^\ast_\ptr t) = 
  \sum_{\kappa\in K} P(\kappa)$.
\end{proposition}

\begin{proof}
By definition, $P(s\to^\ast_\ptr t) 
= \sum_{\cW\in \mathbb{W}_\ptr} 
P(s\to^\ast_\ptr t\mid \cW) \cdot 
P(\cW)$.
We know that $K$ is a covering set of explanations by
Corollary~\ref{cor:covering}. Hence, we have
$\sum_{\cW\in \mathbb{W}_\ptr} 
P(s\to^\ast_\ptr t\mid \cW) \cdot 
P(\cW)
= \sum_{\cW\in \cW_K} P(\cW)$.
Moreover, since $K$ is mutually incompatible,
by Lemma~\ref{lemma:incom}, we have 
$\sum_{\cW\in \cW_K} P(\cW)
= \sum_{\kappa\in K} P(\kappa)$.
\end{proof}
For instance, let $\kappa_1$ and $\kappa_2$ be the 
explanations computed in Example~\ref{ex:sumprob}, i.e.,
$\kappa_1 = \{(R_1,1),(R_2,2)\}$,
$\kappa_2 = \{(R_1,2),(R_2,1)\}$. Then, we have
$P(\kappa_1) = \prob(R_1,1)*\prob(R_2,2) = 0.5*0.4 = 0.2$
and 
$P(\kappa_2) = \prob(R_1,2)*\prob(R_2,1) = 0.5*0.6 = 0.3$.
Since this set of explanations is covering and
mutually incompatible, we have 
$P(\mathsf{switch}(\mathsf{main})
\to^\ast \mathsf{t2}(\mathsf{heads},\mathsf{tails}))
= P(\kappa_1)+P(\kappa_2) = 0.2+0.3=0.5$.

We note that, in Example~\ref{ex:sumprob}, 
we have only considered two 
derivations, although there are actually quite a few more. 
In this case, they all compute one of the two 
explanations shown above (either $\kappa_1$ or $\kappa_2$).
Setting a rewrite strategy that is complete for a certain 
class of TRSs (e.g., leftmost innermost rewriting \cite{Fri85b}) 
may help to minimize the number of derivations to consider.

Unfortunately, 
when a covering set of explanations $K$ for $s\to^\ast t$
is not mutually incompatible,  
the equality 
$P(s\to^\ast t) = \sum_{\kappa\in K} P(\kappa)$ does not generally hold,  
since the sets of worlds that are compatible with 
the explanations 
may overlap, i.e., we can have $\kappa_1,\kappa_2\in K$ 
such that $\cW_{\kappa_1}\cap\cW_{\kappa_2}\neq\emptyset$
and, thus, $P(\kappa_1)+P(\kappa_2)\neq
\sum_{\cW\in\cW_{\kappa_1}\cup\cW_{\kappa_2}} P(\cW)$.

\begin{example}
  Consider again the \ptrs\ of Example~\ref{ex:coins2}, where
  we now add the following two regular rewrite rules:
  \[
  \blue{
  \begin{array}{lrcl}
  R_7: & \mathsf{switch}(\mathsf{t2}(x,\mathsf{heads}))
  & \to & \mathsf{t2}(\mathsf{heads},\mathsf{tails})\\
  R_8: & \mathsf{switch}(\mathsf{t2}(\mathsf{heads},y))
  & \to & \mathsf{t2}(\mathsf{heads},\mathsf{tails})
  \end{array}
  }
  \]
  They represent a sort of shortcut when one coin
  lands on heads. Now, for instance, we have the following
  derivation:
   \[
  \blue{
  \begin{array}{lll}
  \tuple{\mathsf{switch}(\mathsf{main}),\emptyset} & \pto &
  \tuple{\mathsf{switch}(\mathsf{flip2coins}(\mathsf{coin1},  \mathsf{coin2})),\emptyset}\\ 
  & \pto & \tuple{\mathsf{switch}(\mathsf{t2}(\mathsf{coin1},  \mathsf{coin2})),\emptyset}\\
  & \stackrel{0.5}{\pto} & \tuple{\mathsf{switch}(\mathsf{t2}(\mathsf{heads},  \mathsf{coin2})),\{(R_1,1)\}}\\
  & \pto & \tuple{\mathsf{t2}(\mathsf{heads}, \mathsf{tails}),\{(R_1,1)\}}\\  
  \end{array}
  }
  \]
  that computes the explanation $\kappa_3 = \{(R_1,1)\}$
  with $P(\kappa_3) = 0.5$. Here, given $\kappa_1
  = \{(R_1,1),(R_2,2)\}$ computed before, we have
  $\cW_{\kappa_1}\cap\cW_{\kappa_3}\neq \emptyset$.
  Analogously, the derivation:
   \[
  \blue{
  \begin{array}{lll}
  \tuple{\mathsf{switch}(\mathsf{main}),\emptyset} & \pto &
  \tuple{\mathsf{switch}(\mathsf{flip2coins}(\mathsf{coin1},  \mathsf{coin2})),\emptyset}\\ 
  & \pto & \tuple{\mathsf{switch}(\mathsf{t2}(\mathsf{coin1},  \mathsf{coin2})),\emptyset}\\
  & \stackrel{0.6}{\pto} & \tuple{\mathsf{switch}(\mathsf{t2}(\mathsf{coin1},  \mathsf{heads})),\{(R_2,1)\}}\\
  & \pto & \tuple{\mathsf{t2}(\mathsf{heads}, \mathsf{tails}),\{(R_1,1)\}}\\  
  \end{array}
  }
  \]
  computes the explanation $\kappa_4 = \{(R_2,1)\}$
  with $P(\kappa_4) = 0.6$. 
  If we add up all their probabilities, we get 
  $P(\kappa_1) + P(\kappa_2) + P(\kappa_3) + P(\kappa_4) 
  = 0.2 + 0.3 + 0.5 + 0.6 = 1.6$, 
  which is obviously incorrect.
\end{example}
As in the literature on probabilistic logic programming
(see, e.g., \cite{dRKT07}), we can compile
the covering
set of explanations to a BDD (or some of its more 
efficient variants).
Consider, for instance, the above set of covering
explanations:
\[
\blue{K = \{\{(R_1,1)\},\{(R_2,1)\},\{(R_1,1),(R_2,2)\},\{(R_1,2),(R_2,1)\}\}}
\]
Here, since probabilistic rules only have two choices,
we can associate a Boolean variable to each rule, so that
$X_1$ represents the first choice in rule $R_1$
 ($\mathsf{heads}$)
while $\neg X_1$ represents the second choice in this rule
($\mathsf{tails}$),
and similarly $X_{2}$ (first choice of rule $R_2$, 
i.e., $\mathsf{heads}$) and $\neg X_{2}$ (second choice of
rule $R_2$, i.e., $\mathsf{tails}$).
Then, the probability of the considered reachability
problem is given by the probability of the following formula:
\[
\blue{
f_K(\mathbf{X}) = X_{1}\vee X_{2}\vee(X_{1}\wedge \neg X_{2})
\vee (\neg X_{1}\wedge X_{2})
}
\]
that is obtained from the set $K$.
In turn, this formula can be represented with the BDD of Figure~\ref{fig:bdd} (a), where edges are labeled with the associated
probability. As usual, a solid line issuing from a node
represents that the variable is true and a dashed line 
that it is false. 
\begin{figure}[t]
\centering
\begin{minipage}{.45\linewidth}
$
\xymatrix@R=-5pt@C=50pt{
%%%1
&& *++[o][F-]{0} \\
%%%2
& *++[o][F-]{X_{2}} \ar@{-->}@/^/[ur]^{0.4} \ar@/_/[dr]^{0.6}\\
%%%3
&& *++[o][F-]{1} \\
%%%4
*++[o][F-]{X_{1}} \ar@{-->}@/^/[uur]^{0.5} \ar@/_/[ddr]^{0.5}\\
%%%5
&& *++[o][F-]{1} \\
%%%6
& *++[o][F-]{X_{2}} \ar@{-->}@/^/[ur]^{0.4} \ar@/_/[dr]^{0.6}\\
%%%7
&& *++[o][F-]{1} \\
}
$
\end{minipage}
~$\Longrightarrow$~~
\begin{minipage}{.4\linewidth}
$
\xymatrix@R=0pt@C=30pt{
& *++[o][F-]{X_{2}} \ar@{-->}@/^/[r]^{0.4} \ar@/_/[ddr]^{0.6} & *++[o][F-]{0} \\
*++[o][F-]{X_{1}} \ar@{-->}@/^/[ur]^{0.5} \ar@/_/[rrd]_{0.5} \\
& & *++[o][F-]{1} \\
}
$
\end{minipage}

\vspace{1ex}

{\small (a) Original BDD \hspace{16ex} (b) Simplified BDD}
\caption{BDD for formula $X_{1}\vee X_{2}\vee(X_{1}\wedge \neg X_{2})
\vee (\neg X_{1}\wedge X_{2})$} \label{fig:bdd}
\end{figure}
This BDD can be simplified as shown in Figure~\ref{fig:bdd}~(b)
by avoiding repeated nodes and deleting those nodes 
where both edges reach the same value (a so-called 
Reduced Ordered BDD). 

The probability of the formula being true
can now be computed as the sum of
the probabilities of the paths from $X_1$ to $1$,
where the probability of a path is given by the
product of the probabilities of its edges.
Therefore, $P(\mathsf{switch}(\mathsf{main})
\to^\ast \mathsf{t2}(\mathsf{heads},\mathsf{tails})) 
= 0.5+0.5*0.6=0.8$.

This technique can also be generalized to arbitrary rules with
more than two choices using 
\emph{multivalued} decision diagrams (MDDs, \cite{SKMB90}).
Computing the probability from a covering set
of explanations is out of the scope of this work.
Nevertheless, the process is 
essentially identical to that in the literature 
on (exact) inference in probabilistic logic programming; 
thus we refer the interested reader to \cite[Chapter 8]{Rig18}
and references therein.

Finally, for those problems where exact inference is not 
viable, we can consider \emph{approximate} inference.
We distinguish two possible approaches:
\begin{itemize}
\item The first technique adapts \cite{dRKT07,KCRDR08} and
is based on using iterative deepening
when computing a covering set of explanations. 
Given a reachability problem $s\to^\ast t$ and a given 
\emph{depth} $n$, the following two sets of
explanations are produced:\footnote{We 
use $\tuple{s,\kappa} \pto^n \tuple{t,\kappa'}$ 
to denote that $\tuple{s,\kappa}$ is reduced to
$\tuple{t,\kappa'}$ in exactly $n$ steps.}
\[
\begin{array}{l}
\mathbb{E}_l(s\to^\ast_\ptr t) = \{ \kappa \mid \tuple{s,\emptyset} \pto^{m}_\ptr \tuple{t,\kappa},~ m\sleq n\} \\
\mathbb{E}_u(s\to^\ast_\ptr t) = 
\mathbb{E}_l(s\to^\ast_\ptr t) \cup 
\{ \kappa \mid \tuple{s,\emptyset} \pto^{n}_\ptr 
\tuple{t',\kappa}~\wedge~t\neq t'\} \\
\end{array}
\]
Here, $\mathbb{E}_l$ computes the explanations 
associated with those derivations of length less than 
or equal to $n$ that reach the term $t$, 
while $\mathbb{E}_u$ includes, in addition to the 
previous explanations, those associated with derivations 
that are still incomplete after $n$ reduction steps.
By computing the probability associated to each set
(using the same techniques from exact inference seen before),
one gets an interval $[p_u,p_l]$ for the probability.
If the accuracy is deemed adequate, the process ends. 
Otherwise, one can increase the depth $n$ and repeat 
the process.

\item Another technique is based on Monte Carlo sampling
(see, e.g., \cite[Chapter 10]{Rig18}). 
In this case, the idea is to generate a sample of 
possible worlds by considering random choices for the 
(ground instances of the) probabilistic rewrite rules. 
Then, we check whether $s \to^\ast t$ can be proved 
in each generated world. The probability is then 
obtained as a fraction of the worlds where $s \to^\ast t$ 
could be proved. Here, the number of worlds determines 
the confidence interval, so the more worlds considered, 
the more accurate the computed probability will be.
\end{itemize}

%%%%%%%%%%%%%%%%%%%%%%%%%%%%%%%%%%%%%%%%%%%%%%%%%%%%%%%%%%%%
\section{Modeling Problems with Uncertainty} 

In this section, we explore the expressive power of 
\ptrs s for modeling problems with uncertainty and 
propose some extensions.

\begin{example}  \label{ex:covid}
Consider the following \ptrs\ which specifies the level of
protection (against covid-19) depending on the vaccination 
status and the mask worn:
\[
\blue{\begin{array}{l@{~~}rcl@{~~~}lrcl}
R_1: & \mathsf{status}(\mathsf{senior}) 
& \to & 0.9\cons \mathsf{vacc};0.1\cons\mathsf{non\_vacc}\\
R_2: & \mathsf{status}(\mathsf{adult}) 
& \to & 0.5\cons \mathsf{vacc};0.5\cons\mathsf{non\_vacc}\\
R_3: & \mathsf{status}(\mathsf{young}) 
& \to & 0.2\cons \mathsf{vacc};0.8\cons\mathsf{non\_vacc}\\[1ex]
R_4: & \mathsf{mask} & \to &
0.2\cons \mathsf{ffp2};
0.1\cons \mathsf{surgical};
0.7\cons \mathsf{none}\\[1ex]
R_5: & \mathsf{protection}(x) 
& \to & \mathsf{lookup}(\mathsf{status}(x),\mathsf{mask})\\[1ex]
R_6: & \mathsf{lookup}(\mathsf{vacc},\mathsf{ffp2}) 
& \to & \mathsf{strong}\\
R_7: & \mathsf{lookup}(\mathsf{vacc},\mathsf{surgical}) 
& \to & \mathsf{strong}\\
R_8: & \mathsf{lookup}(\mathsf{non\_vacc},\mathsf{ffp2}) 
& \to & \mathsf{strong}\\
R_9: & \mathsf{lookup}(\mathsf{non\_vacc},\mathsf{surgical}) 
& \to & \mathsf{weak}\\
R_{10}: & \mathsf{lookup}(x,\mathsf{none}) 
& \to & \mathsf{no}\\
\end{array}
}
\]
Note that vaccination status depends on a person's age 
while wearing a mask is considered a generic property
which is independent of age in this example.
Here, we may be interested in knowing the probability 
that a senior person has strong protection against covid, 
i.e., $P(\mathsf{protection}(\mathsf{senior})\to^\ast \mathsf{strong})$.

\begin{figure}[t]
\[
\blue{
\begin{array}{lll}
\tuple{\mathsf{protection}(\mathsf{senior}),\emptyset} & \pto & 
\tuple{\mathsf{lookup}(\mathsf{status}(\mathsf{senior}),\mathsf{mask}),\emptyset}\\
&\stackrel{0.9}{\pto} & \tuple{\mathsf{lookup}(\mathsf{vacc},\mathsf{mask}),\{(R_1,1)\}}\\
&\stackrel{0.2}{\pto} & \tuple{\mathsf{lookup}(\mathsf{vacc},\mathsf{ffp2}),\{(R_1,1),(R_4,1)\}}\\
&\pto & \tuple{\mathsf{strong},\{(R_1,1),(R_4,1)\}}\\[1ex]

\tuple{\mathsf{protection}(\mathsf{senior}),\emptyset} & \pto & 
\tuple{\mathsf{lookup}(\mathsf{status}(\mathsf{senior}),\mathsf{mask}),\emptyset}\\
&\stackrel{0.9}{\pto} & \tuple{\mathsf{lookup}(\mathsf{vacc},\mathsf{mask}),\{(R_1,1)\}}\\
&\stackrel{0.1}{\pto} & \tuple{\mathsf{lookup}(\mathsf{vacc},\mathsf{surgical}),\{(R_1,1),(R_4,2)\}}\\
&\pto & \tuple{\mathsf{strong},\{(R_1,1),(R_4,2)\}}\\[1ex]

\tuple{\mathsf{protection}(\mathsf{senior}),\emptyset} & \pto & 
\tuple{\mathsf{lookup}(\mathsf{status}(\mathsf{senior}),\mathsf{mask}),\emptyset}\\
&\stackrel{0.1}{\pto} & \tuple{\mathsf{lookup}(\mathsf{no\_vacc},\mathsf{mask}),\{(R_1,2)\}}\\
&\stackrel{0.2}{\pto} & \tuple{\mathsf{lookup}(\mathsf{no\_vacc},\mathsf{ffp2}),\{(R_1,2),(R_4,1)\}}\\
&\pto & \tuple{\mathsf{strong},\{(R_1,2),(R_4,1)\}}\\[1ex]
\end{array}
}
\]
\caption{Probabilistic derivations for 
$\mathsf{protection}(\mathsf{senior})
\to^\ast \mathsf{strong}$} \label{fig:covid}
\end{figure}

For this purpose, we get a set of explanations $K$ from the
probabilistic rewrite derivations shown in Figure~\ref{fig:covid}.
The computed set of explanations is thus as follows:
\[
\blue{
K = \{\underbrace{\{(R_1,1),(R_4,1)\}}_{\kappa_1},~
\underbrace{\{(R_1,1),(R_4,2)\}}_{\kappa_2},~
\underbrace{\{(R_1,2),(R_4,1)\}}_{\kappa_3}\}
}
\]
with $P(\kappa_1) = 0.18$, $P(\kappa_2) = 0.09$ and
$P(\kappa_3) = 0.02$. In this case, the explanations in $K$
are already mutually incompatible and, thus, we have
$P(\mathsf{protection}(\mathsf{senior})\to^\ast \mathsf{strong})
= P(\kappa_1)+P(\kappa_2)+P(\kappa_3) = 0.29$.
\end{example}

%%%%%%%%%%%%%%%%%%%%%%%%%%%%%%%%%%%%%%%%%%%%%%%
\subsection{Probabilistic Rules with Variables} \label{sec:vars}

Variables in rewrite rules represent a very expressive 
resource. In the previous section, we have only 
considered probabilistic \emph{ground} rules. 
Fortunately, there are 
several ways we can extend the definition 
of probabilistic rewrite rule to also include variables.
First, one can have probabilistic rules with variables 
as long as the domain of such variables is finite. 
Here, probabilistic rules with variables are seen 
only as a compact representation of a (finite) set of ground rules. 
These rules can then be replaced by the 
corresponding instances in a \emph{grounding} preprocessing stage. For this purpose, introducing types would
be useful (as in simply typed term rewriting \cite{Yam01}).

Here, we extend the notion of
atomic choice to a triple $(R,\theta,i)$ where $R,i$ are
the rule and the choice (as before), and the ground
substitution $\theta$ identifies the
considered (ground) instance of rule $R$. The associated
probability is still $\prob(R,i)$ since all
instances have the same choices and probabilities.
In this case, we can say that each ground instance of each 
probabilistic rule represents a random variable of the model.

\begin{example} \label{ex:covid2}
Consider again the \ptrs\ of Example~\ref{ex:covid}. 
Now, however, we consider that function $\mathsf{mask}$
takes the age group ($\mathsf{senior}$, $\mathsf{adult}$,
and $\mathsf{young}$) as argument.
Thus, we have the following \ptrs:
\[
\blue{\begin{array}{l@{~~}rcl@{~~~}lrcl}
R_1: & \mathsf{status}(\mathsf{senior}) 
& \to & 0.9\cons \mathsf{vacc};0.1\cons\mathsf{non\_vacc}\\
R_2: & \mathsf{status}(\mathsf{adult}) 
& \to & 0.5\cons \mathsf{vacc};0.5\cons\mathsf{non\_vacc}\\
R_3: & \mathsf{status}(\mathsf{young}) 
& \to & 0.2\cons \mathsf{vacc};0.8\cons\mathsf{non\_vacc}\\[1ex]
R_4: & \mathsf{mask}(x) & \to &
0.2\cons \mathsf{ffp2};
0.1\cons \mathsf{surgical};
0.7\cons \mathsf{none}\\[1ex]
R_5: & \mathsf{protection}(x) 
& \to & \mathsf{lookup}(\mathsf{status}(x),\mathsf{mask}(x))\\[1ex]
R_6: & \mathsf{lookup}(\mathsf{vacc},\mathsf{ffp2}) 
& \to & \mathsf{strong}\\
R_7: & \mathsf{lookup}(\mathsf{vacc},\mathsf{surgical}) 
& \to & \mathsf{strong}\\
R_8: & \mathsf{lookup}(\mathsf{non\_vacc},\mathsf{ffp2}) 
& \to & \mathsf{strong}\\
R_9: & \mathsf{lookup}(\mathsf{non\_vacc},\mathsf{surgical}) 
& \to & \mathsf{weak}\\
R_{10}: & \mathsf{lookup}(x,\mathsf{none}) 
& \to & \mathsf{no}\\
\end{array}
}
\]
Here, we assume that, according to the type of 
function $\mathsf{mask}$, it can only take
an argument from
$\{\mathsf{senior},\mathsf{adult},\mathsf{young}\}$.
Then, rule $R_4$ will be replaced in a grounding stage by
\[
\blue{
\begin{array}{rcl}
%(R_{41}) & 
\mathsf{mask}(\mathsf{senior})  & \to &
0.2\cons \mathsf{ffp2};
0.1\cons \mathsf{surgical};
0.7\cons \mathsf{none}\\
%(R_{42}) & 
\mathsf{mask}(\mathsf{adult})  & \to &
0.2\cons \mathsf{ffp2};
0.1\cons \mathsf{surgical};
0.7\cons \mathsf{none}\\
%(R_{43}) & 
\mathsf{mask}(\mathsf{young}) & \to &
0.2\cons \mathsf{ffp2};
0.1\cons \mathsf{surgical};
0.7\cons \mathsf{none}
\end{array}
}
\]
Note that, in contrast to Example~\ref{ex:covid} where
everyone wore the same type of mask, now we can have
different choices for each age group.
\end{example}
In principle, the case above does not imply any 
relevant change with 
respect to the theory seen in Section~\ref{sec:ptrs},
since the same conditions apply after the grounding stage.

Let us now consider the general case in which variables 
can be bound to infinitely many different terms and, 
thus, it makes no sense to consider a grounding stage.
In this case, a \ptrs\ encodes a possibly 
infinite number of random variables (as many as ground 
instances of the probabilistic rules exist).
Therefore, the probability 
of a world cannot be defined as the product of the
choices in a selection,
$P(\cW_\sel) = \prod_{(R,\theta,i)\in \sel} \prob(R,i)$,
since it would converge to zero whenever a selection
includes infinitely many choices.

Following \cite{Poo00}, in order to overcome the 
above problem, one can introduce a measure over
(possibly infinite) \emph{sets} of possible worlds:
\begin{itemize}
\item Given a finite set of (finite) composite choices 
$K$, the probability of the (possibly infinite) 
set of worlds $\cW_K$ is given by $\mu(K)$, which
is defined in terms of a set $K'$ of mutually 
incompatible composite choices that is 
\emph{equivalent} to $K$, i.e., that represents the same 
(possibly infinite) set of worlds,
$\cW_K = \cW_{K'}$. 
Then, $\mu(K) = 
\sum_{\kappa\in K'} P(\kappa) =
\sum_{\kappa\in K'} \prod_{(R,\theta,i)\in\kappa} \prob(R,i)$.
Such a set $K'$ can always be obtained
from $K$ by \emph{splitting} \cite{Poo00}, a transformation
that applies the following
two operations repeatedly until a mutually incompatible
set is obtained:
\begin{enumerate}
\item If $\kappa_1,\kappa_2\in K$ with 
$\kappa_1\subset\kappa_2$, then remove $\kappa_2$ from $K$.
\item If $\kappa_1,\kappa_2\in K$ with 
$\kappa_1\cup\kappa_2$ consistent, find some probabilistic 
rule $R$
which only occurs in $\kappa_1$ and replace $\kappa_2$
in $K$ by the sets $\kappa_2\cup\{(R,1)\}$,
\ldots, $\kappa_2\cup\{(R,n)\}$ if $R$ has $n$ possible choices.
\end{enumerate}

\item Then, given a reachability problem, $s\to^\ast_\ptr t$
and a finite set of (finite) covering explanations $K$,
we let $P(s\to^\ast_\ptr t) = \mu(K)$.
\end{itemize}
Therefore, introducing probabilistic rules with variables
over possibly infinite domains
is not a problem as long as we can obtain a finite covering set
of (finite) explanations. 
In particular, we would need $\theta$ to be ground in
every atomic choice $(R,\theta,i)$. For this purpose, 
we could require, e.g., that the term to be reduced is 
ground (and, consequently, all the terms in its derivations 
since the considered \ptrs s have no extra variables), 
which is a reasonable requirement in practice.

\begin{example} \label{ex:covid3}
  Consider the \ptrs\ of Example~\ref{ex:covid2}
  (with no grounding stage). For the
  reachability problem $\mathsf{protection}(\mathsf{senior})\to^\ast
  \mathsf{strong}$, we can generate the following probabilistic
  rewrite derivations:
\[
\blue{
\begin{array}{lll}
\tuple{\mathsf{protection}(\mathsf{senior}),\emptyset} & \pto & 
\tuple{\mathsf{lookup}(\mathsf{status}(\mathsf{senior}),\mathsf{mask}(\mathsf{senior})),\emptyset}\\
&\stackrel{0.9}{\pto} & \tuple{\mathsf{lookup}(\mathsf{vacc},\mathsf{mask}(\mathsf{senior})),\{(R_1,\id,1)\}}\\
&\stackrel{0.2}{\pto} & \tuple{\mathsf{lookup}(\mathsf{vacc},\mathsf{ffp2}),\{(R_1,\id,1),(R_4,\{x/\mathsf{senior}\},1)\}}\\
&\pto & \tuple{\mathsf{strong},\{(R_1,\id,1),(R_4,\{x/\mathsf{senior}\},1)\}}\\[1ex]

\tuple{\mathsf{protection}(\mathsf{senior}),\emptyset} & \pto & 
\tuple{\mathsf{lookup}(\mathsf{status}(\mathsf{senior}),\mathsf{mask}(\mathsf{senior})),\emptyset}\\
&\stackrel{0.9}{\pto} & \tuple{\mathsf{lookup}(\mathsf{vacc},\mathsf{mask}(\mathsf{senior})),\{(R_1,\id,1)\}}\\
&\stackrel{0.1}{\pto} & \tuple{\mathsf{lookup}(\mathsf{vacc},\mathsf{surgical}),\{(R_1,\id,1),(R_4,\{x/\mathsf{senior}\},2)\}}\\
&\pto & \tuple{\mathsf{strong},\{(R_1,\id,1),(R_4,\{x/\mathsf{senior}\},2)\}}\\[1ex]

\tuple{\mathsf{protection}(\mathsf{senior}),\emptyset} & \pto & 
\tuple{\mathsf{lookup}(\mathsf{status}(\mathsf{senior}),\mathsf{mask}(\mathsf{senior})),\emptyset}\\
&\stackrel{0.1}{\pto} & \tuple{\mathsf{lookup}(\mathsf{no\_vacc},\mathsf{mask}(\mathsf{senior})),\{(R_1,\id,2)\}}\\
&\stackrel{0.2}{\pto} & \tuple{\mathsf{lookup}(\mathsf{no\_vacc},\mathsf{ffp2}),\{(R_1,\id,2),(R_4,\{x/\mathsf{senior}\},1)\}}\\
&\pto & \tuple{\mathsf{strong},\{(R_1,\id,2),(R_4,\{x/\mathsf{senior}\},1)\}}\\[1ex]
\end{array}
}
\]
The computed set $K = \{\kappa_1,\kappa_2,\kappa_3\}$
of explanations is now as follows:
\[
\blue{
\begin{array}{lll}
\kappa_1 & = & \{(R_1,\id,1),(R_4,\{x/\mathsf{senior}\},1)\},
~~\mbox{with}~P(\kappa_1) = 0.9*0.2=0.18\\
\kappa_2 & = & \{(R_1,\id,1),(R_4,\{x/\mathsf{senior}\},2)\},
~~\mbox{with}~P(\kappa_2) = 0.9*0.1=0.09\\
\kappa_3 & = & \{(R_1,\id,2),(R_4,\{x/\mathsf{senior}\},1)\},
~~\mbox{with}~P(\kappa_3) = 0.1*0.2=0.02\\
\end{array}
}
\]
and, thus, the probability of 
$\mathsf{protection}(\mathsf{senior})\to^\ast \mathsf{strong}$ 
is the same as before, i.e., 
$P(\mathsf{protection}(\mathsf{senior})\to^\ast \mathsf{strong}) = P(\kappa_1)+P(\kappa_2)+P(\kappa_3) = 0.29$.
\end{example}

%%%%%%%%%%%%%%%%%%%%%%%%%%%%%%%%%%%%%%%%%%%%%%%%%%%%%%%%%%%
\subsection{Conditional \ptrs s}

Another interesting extension from the point of view 
of expressive power consists of considering conditional 
rewrite rules. For this purpose, let us first briefly 
introduce \emph{conditional} term rewrite
systems (CTRSs); namely oriented 3-CTRSs, i.e., CTRSs where extra
variables are allowed as long as $\Var(r)\subseteq\Var(l)\cup\Var(C)$
for any rule $l\to r\Leftarrow C$ \cite{MH94}.
In \emph{oriented} CTRSs, a conditional rule $l\to r \Leftarrow C$ has the
form $l\to r \Leftarrow s_1\tto t_1,\ldots,s_m\tto t_m$, where each
oriented equation $s_i\tto t_i$ is interpreted as reachability.

For a CTRS $\cR$, the associated rewrite relation $\to_\cR$ is defined
as the smallest binary relation satisfying the following: given
terms $s,t\in\cT(\cF)$, we have $s \to_\cR t$ iff there exist
a position $p$ in $s$, a rewrite rule $l \to r\Leftarrow 
s_1\tto t_1,\ldots,s_m\tto t_m\in \cR$, and 
a substitution $\sigma$ such that $s|_p
= l\sigma$, $s_i\sigma \to^\ast_\cR t_i\sigma$ for all $i=1,\ldots,m$,
and $t = s[r\sigma]_p$.

The extension of \ptrs s and probabilistic rewriting 
to the conditional case is very natural. 
A probabilistic \emph{conditional} rule has now the form
\[
\blue{
R:l \to p_1\cons r_1;\ldots;p_n\cons r_n
\Leftarrow s_1\tto t_1,\ldots,s_m\tto t_m
}
\]
A \emph{probabilistic conditional term rewrite 
system} (\pctrs) is a disjoint union of a set of probabilistic
conditional rules and a set of 
regular conditional rewrite rules. 
A \emph{world} is then obtained from a selection $\sel$ as
in the unconditional case, i.e., by choosing one
right-hand side per probabilistic conditional 
rule:\footnote{Here, we assume that probabilistic conditional
rules are ground for simplicity, but the same considerations
of Section~\ref{sec:vars} regarding variables apply.}
\[
\blue{\cW_\sel
= \{ l \to r_i \Leftarrow C \mid
R:(l\to p_1\cons r_1;\ldots;p_n\cons r_n \leftarrow
C)\in\ptr_p\wedge
(R,i)\in \sel\} \cup \ptr_r}
\]
where $C$ is a (possibly empty) sequence of oriented equations
of the form $s_i\tto t_i$, $i=1,\ldots,m$, $m\geq 0$.
Definition~\ref{def:lifting} applies to 
\pctrs s too: a property $P$ holds for \pctrs\ $\ptr$
  if $P$ holds for each world $\cW\in\mathbb{W}_\ptr$.

Probabilistic conditional rewriting can then be defined
as follows:

\begin{definition}[probabilistic conditional rewriting]
Let $s,t$ be terms and $\kappa,\kappa'$ be composite choices.
A \pctrs\ $\ptr$ induces a \emph{probabilistic conditional 
rewrite} relation $\pto_\ptr$ 
where $\tuple{s,\kappa} \pto_\ptr \tuple{t,\kappa'}$
if either
\begin{itemize}
\item there is a position
$\pi$ in $s$, a regular rewrite rule $l \to r \Leftarrow
s_1\tto t_1,\ldots,s_m\tto t_m\in \ptr$, and a 
substitution $\sigma$ such that $s|_\pi = l\sigma$, 
$\tuple{s_i\sigma,\kappa} \pto^\ast_\ptr 
\tuple{t_i\sigma,\kappa_i}$ for all $i=1,\ldots,m$,
$t = s[r\sigma]_\pi$, and $\kappa'=\kappa_1\cup\ldots\cup\kappa_m$
is consistent, or

\item there is a position
$\pi$ in $s$, a probabilistic rewrite rule 
$R: l \to p_1\cons r_1;\ldots;p_n\cons r_n \Leftarrow
s_1\tto t_1,\ldots,s_m\tto t_m\in \ptr$, and a 
substitution $\sigma$ such that $s|_\pi = l\sigma$, 
$\tuple{s_i\sigma,\kappa} \pto^\ast_\ptr 
\tuple{t_i\sigma,\kappa_i}$ for all $i=1,\ldots,m$,
$t = s[r_j\sigma]_\pi$, $j\in\{1,\ldots,n\}$, and 
$\kappa'=\kappa_1\cup\ldots\cup\kappa_m\cup\{(R,j)\}$
is consistent.
\end{itemize}
\end{definition}
We label probabilistic conditional steps with a 
probability, similarly to the unconditional case:
given 
$\tuple{s,\kappa} \stackrel{p}{\pto} 
\tuple{t,\kappa'}$,
we have $p=P(\kappa'')$ if 
$\kappa'\setminus\kappa = \kappa''$,
and $p=1$ otherwise. 
The only difference is that, now, we can add
more than one atomic choice in a single step
(because of the evaluation of the conditions).
Probabilities are then lifted to derivations by
computing the product of the probabilities of their steps.

\begin{example} \label{ex:alarm}
  Consider the following \pctrs :
\[
\blue{\begin{array}{l@{~~}rcl@{~~~}lrcl}
R_1: & \mathsf{crime} & \to & 0.2\cons\mathsf{burglary}; 0.8\cons\mathsf{other}\\
R_2: & \mathsf{earthquake} & \to & 0.2\cons\mathsf{strong}; 
0.5\cons\mathsf{mild}; 0.3\cons\mathsf{none}\\[1ex]

R_3: & \mathsf{alarm} & \to & 0.9\cons\mathsf{ring}
%; 0.1\cons\mathsf{off}
\Leftarrow \mathsf{crime} \tto \mathsf{burglary}, \mathsf{earthquake} \tto \mathsf{strong}\\

R_4: & \mathsf{alarm} & \to & 0.8\cons\mathsf{ring}
%; 0.2\cons\mathsf{off} 
\Leftarrow \mathsf{crime} \tto \mathsf{burglary}, \mathsf{earthquake} \tto \mathsf{mild}\\

R_5: & \mathsf{alarm} & \to & 0.7\cons\mathsf{ring}
%; 0.3\cons\mathsf{off} 
\Leftarrow \mathsf{crime} \tto \mathsf{burglary}, \mathsf{earthquake} \tto \mathsf{none}\\

R_6: & \mathsf{alarm} & \to & 0.3\cons\mathsf{ring}
%; 0.7\cons\mathsf{off} 
\Leftarrow \mathsf{crime} \tto \mathsf{other}, \mathsf{earthquake} \tto \mathsf{strong}\\

R_7: & \mathsf{alarm} & \to & 0.1\cons\mathsf{ring}
%; 0.9\cons\mathsf{off} 
\Leftarrow \mathsf{crime} \tto \mathsf{other}, \mathsf{earthquake} \tto \mathsf{mild}\\
\end{array}
}
\]
We note that ``$\mathsf{other}$'' in rule $R_1$ represents
either other types of crimes and no crime at all. 
Also, notice that the right-hand sides of rules 
$R_3-R_7$ have an implicit choice (i.e., $\bot$) representing
that the alarm does not ring.

In this case, we are interested in computing the probability that 
the alarm rings, i.e., that $\mathsf{alarm}$ can be 
reduced to $\mathsf{ring}$. Here, we have the 
probabilistic conditional rewrite derivations
shown in Figure~\ref{fig:alarm}.
\begin{figure}[t]
\[
\blue{
\begin{array}{rlll}
D_1= & \tuple{\mathsf{alarm},\emptyset}
& \stackrel{0.036}{\pto} & \tuple{\mathsf{ring},
  \{(R_1,1),(R_2,1),(R_3,1)\}}\\
&&& \mbox{since} ~ \tuple{\mathsf{crime},\emptyset}
 \stackrel{0.2}{\pto}  \tuple{\mathsf{burglary},\{(R_1,1)\}}\\ 
&&& \mbox{and} ~ \tuple{\mathsf{earthquake},\emptyset}
 \stackrel{0.2}{\pto}  \tuple{\mathsf{strong},\{(R_2,1)\}}\\[1ex]

D_2= & \tuple{\mathsf{alarm},\emptyset}
& \stackrel{0.08}{\pto} & \tuple{\mathsf{ring},
  \{(R_1,1),(R_2,2),(R_4,1)\}}\\
&&& \mbox{since} ~ \tuple{\mathsf{crime},\emptyset}
 \stackrel{0.2}{\pto}  \tuple{\mathsf{burglary},\{(R_1,1)\}}\\ 
&&& \mbox{and} ~ \tuple{\mathsf{earthquake},\emptyset}
 \stackrel{0.5}{\pto}  \tuple{\mathsf{mild},\{(R_2,2)\}}\\[1ex]

D_3= & \tuple{\mathsf{alarm},\emptyset}
& \stackrel{0.042}{\pto} & \tuple{\mathsf{ring},
  \{(R_1,1),(R_2,3),(R_5,1)\}}\\
&&& \mbox{since} ~ \tuple{\mathsf{crime},\emptyset}
 \stackrel{0.2}{\pto}  \tuple{\mathsf{burglary},\{(R_1,1)\}}\\ 
&&& \mbox{and} ~ \tuple{\mathsf{earthquake},\emptyset}
 \stackrel{0.3}{\pto}  \tuple{\mathsf{none},\{(R_2,3)\}}\\[1ex]

D_4= & \tuple{\mathsf{alarm},\emptyset}
& \stackrel{0.048}{\pto} & \tuple{\mathsf{ring},
  \{(R_1,2),(R_2,1),(R_6,1)\}}\\
&&& \mbox{since} ~ \tuple{\mathsf{crime},\emptyset}
 \stackrel{0.8}{\pto}  \tuple{\mathsf{other},\{(R_1,2)\}}\\ 
&&& \mbox{and} ~ \tuple{\mathsf{earthquake},\emptyset}
 \stackrel{0.2}{\pto}  \tuple{\mathsf{strong},\{(R_2,1)\}}\\[1ex]

D_5= & \tuple{\mathsf{alarm},\emptyset}
& \stackrel{0.04}{\pto} & \tuple{\mathsf{ring},
  \{(R_1,2),(R_2,2),(R_7,1)\}}\\
&&& \mbox{since} ~ \tuple{\mathsf{crime},\emptyset}
 \stackrel{0.8}{\pto}  \tuple{\mathsf{other},\{(R_1,2)\}}\\ 
&&& \mbox{and} ~ \tuple{\mathsf{earthquake},\emptyset}
 \stackrel{0.5}{\pto}  \tuple{\mathsf{mild},\{(R_2,2)\}}\\
\end{array}
}
\]
\caption{Probabilistic derivations for
$\mathsf{alarm} \to^\ast \mathsf{ring}$} \label{fig:alarm}
\end{figure}
The probability labeling the first step of derivation
$D_1$ is obtained from 
the product  $\prob(R_1,1)*\prob(R_2,1)*\prob(R_3,1) =
0.2*0.2*0.9 = 0.036$, and analogously for the remaining 
derivations.

Let $K = \{\kappa_1,\ldots,\kappa_5\}$ 
be the computed set of explanations, where
$k_i$ is the explanation computed by derivation $D_i$,
$i = 1,\ldots,5$.
Then, we have $P(\kappa_1) = 0.036$,
$P(\kappa_2) = 0.08$, $P(\kappa_3) = 0.042$, $P(\kappa_4) = 0.048$,
and $P(\kappa_5)=0.04$. Hence, the most likely
explanation for the alarm to ring is $\kappa_2$, i.e.,
there was a burglary and, in parallel, a mild earthquake.
The overall probability that the alarm rings is
$P(\mathsf{alarm}\to^\ast\mathsf{ring}) = 
P(\kappa_1) + P(\kappa_2) + P(\kappa_3) + P(\kappa_4)
+ P(\kappa_5) = 0.246$ since $K$ is not only covering
but also mutually incompatible.
\end{example}
Similarly to the case of probabilistic logic programming, 
we can also encode Bayesian networks \cite{Pearl88} 
using \pctrs s. Basically, each random variable 
is encoded as a Boolean function and each row of its
conditional probability table is encoded
as a conditional rule as follows:

\begin{example}\label{ex:bayesian}
\begin{figure}[t]
\centering

\includegraphics[scale=0.8]{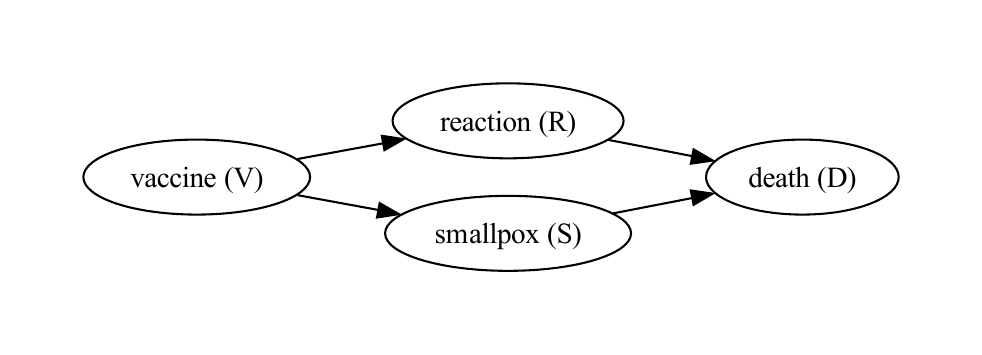}

\begin{tabular}{|l|l|l|}\hline
V & t & f \\\hline
& 0.8 & 0.2 \\ \hline
\end{tabular} 
~~~
\begin{tabular}{|l|l|l|}\hline
R & t & f \\\hline
$V$=t& 0.05 & 0.95 \\ \hline
$V$=f& 0 & 1.0 \\ \hline
\end{tabular}
~~~
\begin{tabular}{|l|l|l|}\hline
S & t & f \\\hline
$V$=t& 0.01 & 0.99 \\ \hline
$V$=f& 0.3 & 0.7 \\ \hline
\end{tabular}
\\[3ex]
\begin{tabular}{|l|l|l|}\hline
D & t & f \\\hline
$R$=t,$S$=t & 0.2 & 0.8 \\ \hline
$R$=t,$S$=f & 0.01 & 0.99\\ \hline
$R$=f,$S$=t & 0.3 & 0.7 \\ \hline
$R$=f,$S$=f & 0 & 1.0 \\ \hline
\end{tabular}

\caption{Bayesian network modeling the probability
of dying from smallpox} \label{fig:pox}
\end{figure}
  Consider the Bayesian network shown in Figure~\ref{fig:pox}
  (freely adapted from \cite{Pearl18}).
  It can be encoded using the following \pctrs:
  \[
  \blue{\begin{array}{l@{~~}rcl@{~~~}lrcl}
  R_1: & \mathsf{vaccine} & \to & 0.8\cons\mathsf{true}; 0.2\cons\mathsf{false}\\[1ex]
  
  R_2: & \mathsf{reaction} & \to & 0.05\cons\mathsf{true}; 0.95\cons\mathsf{false} \Leftarrow \mathsf{vaccine}\tto\mathsf{true}\\
  R_3: & \mathsf{reaction} & \to & 0\cons\mathsf{true}; 1\cons\mathsf{false} \Leftarrow \mathsf{vaccine}\tto\mathsf{false}\\[1ex]

  R_4: & \mathsf{smallpox} & \to & 0.01\cons\mathsf{true}; 0.99\cons\mathsf{false} \Leftarrow \mathsf{vaccine}\tto\mathsf{true}\\
  R_5: & \mathsf{smallpox} & \to & 0.3\cons\mathsf{true}; 0.7\cons\mathsf{false} \Leftarrow \mathsf{vaccine}\tto\mathsf{false}\\[1ex]
  
  R_6: & \mathsf{death} & \to & 0.2\cons\mathsf{true}; 0.8\cons\mathsf{false} \Leftarrow \mathsf{reaction}\tto\mathsf{true},\mathsf{smallpox}\tto\mathsf{true}\\
  R_7: & \mathsf{death} & \to & 0.01\cons\mathsf{true}; 0.99\cons\mathsf{false} \Leftarrow \mathsf{reaction}\tto\mathsf{true},\mathsf{smallpox}\tto\mathsf{false}\\
  R_8: & \mathsf{death} & \to & 0.3\cons\mathsf{true}; 0.7\cons\mathsf{false} \Leftarrow \mathsf{reaction}\tto\mathsf{false},\mathsf{smallpox}\tto\mathsf{true}\\
  R_9: & \mathsf{death} & \to & 0\cons\mathsf{true}; 1\cons\mathsf{false} \Leftarrow \mathsf{reaction}\tto\mathsf{false},\mathsf{smallpox}\tto\mathsf{false}\\
  \end{array}
  }
  \]  
  Here, we are interested in the probability 
  of death, i.e., $P(\mathsf{death}\to^\ast\mathsf{true})$.
  In this case, 
  there exist many possible derivations, some of them
  shown in Figure~\ref{fig:bayesian}.
  \begin{figure}[t]
  \[
  \blue{
  \begin{array}{rlll}
  D_1 = & \tuple{\mathsf{death},\emptyset}
  & \stackrel{0.00008}{\pto} & \tuple{\mathsf{true},
  \{(R_1,1),(R_2,1),(R_4,1),(R_6,1)\}} \\
  &&& \mbox{since} ~ \tuple{\mathsf{reaction},\emptyset}
 \stackrel{0.04}{\pto}  \tuple{\mathsf{true},\{(R_1,1),(R_2,1)\}}\\ 
 &&& \hspace{5ex}
   \mbox{since} ~ \tuple{\mathsf{vaccine},\emptyset}
 \stackrel{0.8}{\pto}  \tuple{\mathsf{true},\{(R_1,1)\}}\\
  &&& \mbox{and} ~ \tuple{\mathsf{smallpox},\emptyset}
 \stackrel{0.008}{\pto}  \tuple{\mathsf{true},\{(R_1,1),(R_4,1)\}}\\
 &&& \hspace{5ex}
   \mbox{since} ~ \tuple{\mathsf{vaccine},\emptyset}
 \stackrel{0.8}{\pto}  \tuple{\mathsf{true},\{(R_1,1)\}}\\[1ex]
 
  D_2 = & \tuple{\mathsf{death},\emptyset}
  & \stackrel{0.000396}{\pto} & \tuple{\mathsf{true},
  \{(R_1,1),(R_2,1),(R_4,2),(R_7,1)\}} \\
  &&& \mbox{since} ~ \tuple{\mathsf{reaction},\emptyset}
 \stackrel{0.04}{\pto}  \tuple{\mathsf{true},\{(R_1,1),(R_2,1)\}}\\ 
 &&& \hspace{5ex}
   \mbox{since} ~ \tuple{\mathsf{vaccine},\emptyset}
 \stackrel{0.8}{\pto}  \tuple{\mathsf{true},\{(R_1,1)\}}\\
  &&& \mbox{and} ~ \tuple{\mathsf{smallpox},\emptyset}
 \stackrel{0.792}{\pto}  \tuple{\mathsf{false},\{(R_1,1),(R_4,2)\}}\\
 &&& \hspace{5ex}
   \mbox{since} ~ \tuple{\mathsf{vaccine},\emptyset}
 \stackrel{0.8}{\pto}  \tuple{\mathsf{true},\{(R_1,1)\}}\\[1ex]

  \ldots
 \end{array}
  }
  \]
  \caption{Some probabilistic derivations for
  $\mathsf{death}\to^\ast \mathsf{true}$} \label{fig:bayesian}
  \end{figure}
  
  It is worth noting that this encoding works well 
  because, once the choice is fixed for a probabilistic rule, 
  it cannot be changed in the remaining derivation. 
  For example, in derivation $D_1$, if the 
  reduction from $\mathsf{reaction}$ to $\mathsf{true}$    
  (in the condition of rule $R_6$) considers that
  $\mathsf{vaccine}$ reduces to $\mathsf{true}$ in order
  to apply rule $R_2$, then the reduction from
  $\mathsf{smallpox}$ to $\mathsf{true}$ can only
  consider rule $R_4$ (where $\mathsf{vaccine}$ should
  also be reduced to $\mathsf{true}$). 
  Here, rule $R_5$ would no longer be
  applicable because that rule requires $\mathsf{vaccine}$
  to be reduced to $\mathsf{false}$.
  This is a relevant aspect that other approaches---like 
  those of Bournez et al.\ \cite{BK02,BG05,BG06}
  and more recent works based on them, e.g., \cite{ALY20,KC17,KC19,KFG24}---to 
  probabilistic rewriting do not consider; typically, 
  in these works, 
  the same probabilistic rule can be used repeatedly throughout 
  a derivation, selecting different choices in different
  steps.
\end{example}

%%%%%%%%%%%%%%%%%%%%%%%%%%%%%%%%%%%%%%%%%%%%%%%%%%%%%%%%%%%%%%%%
\section{Related Work} \label{sec:relwork}

As mentioned in the Introduction, 
the most popular approach to probabilistic rewriting is
that of Bournez et al.\ \cite{BK02,BG05,BG06}. The notions
of probabilistic abstract reduction system and probabilistic
term rewrite systems have been later refined in \cite{ALY20},
where \emph{multidistributions} are introduced to appropriately 
account for both the probabilistic behavior of each rule 
and the nondeterminism in rule selection. A similar notion
of \ptrs\ is considered in \cite{KC17,KC19,KFG24}. 
Our definition of
\ptrs\ is closely related, but our work has a number of 
significant differences.
First, we partition the rules of a \ptrs\ into
a set of probabilistic rules and
a set of regular rules, and require the probabilistic rules
to be ground. This is required for the distribution
semantics to be well-defined (although an extension to
non-ground probabilistic rules is possible along the
lines of Section~\ref{sec:vars}).

Secondly, the probability of a reduction is defined
in these works as the product of the probabilities of the 
rewrite steps in this reduction. 
This is significantly different in our distribution semantics, 
where we consider the probability of a reduction to 
be the product of the atomic choices computed in this reduction.
In particular, this implies the following key differences:
\begin{itemize}
\item First, once we apply a probabilistic rule 
and choose one of the terms in the right-hand side, 
no other choice can be considered in the remaining steps 
of the same derivation. 
Note that this difference becomes relevant, e.g., to
model problems like the Bayesian network of Example~\ref{ex:bayesian}.

\item On the other hand, if we apply a probabilistic rule 
several times in the same derivation, 
its probability is computed only once. 
\end{itemize}
Finally, while these works focus on studying
properties such as termination or confluence of the \ptrs s,
we emphasize their use for modeling problems
with a certain degree of uncertainty.

Other approaches to defining a probabilistic extension of
rewriting include \cite{BH03}, where the aim is to assign
probabilities (or weights) to (ordinary) rewrite rules,
or an extension of Maude \cite{CDELMMQ02} 
where extra-variables in the right-hand
sides of the rules are instantiated according to some
probability distribution \cite{AMS06}.
All the above works consider different ways of integrating 
probabilities in term rewriting and, in particular, 
different from our proposal.

Finally, we can also find probability extensions in some
related, rule-based formalisms like the $\lambda$-calculus
(see, e.g., \cite{LZ12}, and references therein) and
constraint handling rules \cite{FPW02}.
In particular, the language Church \cite{GMRBT08}
has a memoization mode in which, once a probabilistic 
function returning a certain value is used, 
subsequent calls will always return the same value, 
ignoring all other choices. This mode resembles what 
happens in our distribution semantics, where the 
application of a probabilistic rule in a given reduction 
is considered a one-time event that cannot be changed 
later during the same reduction.
However, Church is a language with a very different 
orientation than our \ptrs s. Assume, e.g., that we 
define in Church a function $\mathsf{coin}$ that can 
return $\mathsf{heads}$ or $\mathsf{tails}$ with equal probability. This means that a call to this function
will randomly return either $\mathsf{heads}$ or
$\mathsf{tails}$ (but only one value). Moreover, if we
call function $\mathsf{coin}$ a sufficiently large number 
of times, we will observe that the obtained values are
uniformly distributed. In other words, Church can be used
for describing stochastic generative processes, so that
the output of a program follows a 
particular distribution. 
In contrast, if we define a probabilistic rule like
$\mathsf{coin} \to 0.5\cons \mathsf{heads}; 0.5\cons\mathsf{tails}$, the evaluation of 
$\mathsf{coin}$ is actually 
deterministic and always 
return \emph{both} possible values, $\mathsf{heads}$ and 
$\mathsf{tails}$, each one with its associated 
probability. In other words, \ptrs s are used to specify
models and perform probabilistic inference.

%%%%%%%%%%%%%%%%%%%%%%%%%%%%%%%%%%%%%%%%%%%%%%%%%%%%%%%%%%%%%%%%%%%%%%%
\section{Concluding Remarks and Future Work} \label{sec:conc}

In this paper, we have presented a new approach to combine
term rewriting and probabilities, resulting in a more 
expressive formalism that can model complex relationships 
and uncertainty.
We have defined a distribution semantics 
inspired to Sato's proposal for logic programs \cite{Sat95},
which forms the basis of 
several probabilistic logic languages (e.g.,  
ProbLog \cite{dRKT07}, 
Logic Programs with Annotated Disjunctions \cite{VVB04}, 
Independent Choice Logic \cite{Poo97} or
PRISM \cite{SK97}). 
We thus think that a distribution 
semantics can also play a significant role 
in the context of term rewriting. 
We have also shown how the probability 
of a reduction can be obtained from a set of 
(covering) explanations.
Finally, we have illustrated the expressive power of our approach
by means of several examples, and outlined some possible
extensions.

As future work, we consider it interesting to determine the 
conditions under which a PTRS/PCTRS will always produce a set 
of mutually incompatible explanations, which would allow us to 
calculate the probability of a reachability problem
very easily as the sum of the 
probabilities of its explanations.
Another useful extension involves the design of 
appropriate probabilistic rewrite strategies 
that prune derivations when the reachability problem is 
guaranteed to be infeasible, e.g., by reusing
existing results from reachability analysis.
This will allow us to address the implementation of a 
software tool to calculate the probability 
of a rechability problem.
Finally, we will also consider the definition of methods
to present the computed explanations 
in a way that is understandable for non-experts, similarly 
to what we have done in \cite{Vid24} for probabilistic 
logic programs.

%\subsubsection*{Acknowledgements.}
%
%I would like to thank the anonymous reviewers 
%for their suggestions to improve this paper.

\bibliographystyle{splncs04}

\end{document}